\documentclass[journal]{IEEEtran}

\usepackage{color,soul}

\makeatletter
\newcount\SOUL@minus 
\makeatother

\usepackage{algorithm}
\usepackage[noend]{algpseudocode}
\usepackage{epstopdf}
\usepackage[utf8]{inputenc}
\usepackage{graphicx,color}
\usepackage{cite}
\usepackage{subcaption}
\usepackage{times}  
\usepackage[hyphens]{url}  
\usepackage{mathtools, nccmath}

\usepackage{multirow}

\usepackage{array}
\usepackage{hyperref}
\usepackage[bottom]{footmisc}
\usepackage{amsmath,amsfonts,amssymb}

\makeatletter
\renewcommand{\Function}[2]{%
  \csname ALG@cmd@\ALG@L @Function\endcsname{#1}{#2}%
  \def\jayden@currentfunction{#1}%
}
\newcommand{\funclabel}[1]{%
  \@bsphack
  \protected@write\@auxout{}{%
    \string\newlabel{#1}{{\jayden@currentfunction}{\thepage}}%
  }%
  \@esphack
}
\makeatother

\begin{document}
\title{Context-Aware Deep Q-Network for Decentralized Cooperative Reconnaissance by a Robotic Swarm}
\author{N.~Mohanty,
        M.~S.~Gadde,
        S.~Sundaram,~\IEEEmembership{Senior~Member,~IEEE},
        N.~Sundararajan,~\IEEEmembership{Life~Fellow,~IEEE}
        and~P.B.~Sujit,~\IEEEmembership{Member,~IEEE}
\thanks{N. Mohanty, M.S. Gadde and S. Sundaram are with the Department
of Aerospace Engineering, IISc, Bangalore,  India. (e-mail: (nishantm, mohitvishnug, vssuresh)$@$iisc.ac.in).}

\thanks{N. Sundararajan is from School of Electrical and Electronic Engineering, NTU, Singapore. (e-mail: ensundara$@$ntu.edu.sg).}%
\thanks{P.B. Sujit is with the Department
of Electrical Engineering and Computer Science, IISER, Bhopal,  India. (e-mail: sujit$@$iiserb.ac.in).}}%

%

\maketitle

\begin{abstract}
One of the crucial problems in robotic swarm-based operation is to search and neutralize heterogeneous targets in an unknown and uncertain environment, without any communication within the swarm. Here, some targets can be neutralized by a single robot, while others need multiple robots in a particular sequence to neutralize them. The complexity in the problem arises due to the scalability and information uncertainty, which restricts the robot's awareness of the swarm and the target distribution. In this paper, this problem is addressed by proposing a novel Context-Aware Deep Q-Network (CA-DQN) framework to obtain communication free cooperation between the robots in the swarm. Each robot maintains an adaptive grid representation of the vicinity with the context information embedded into it to keep the swarm intact while searching and neutralizing the targets. The problem formulation uses a reinforcement learning framework where two Deep Q-Networks (DQNs) handle 'conflict' and 'conflict-free' scenarios separately. The self-play-in-based approach is used to determine the optimal policy for the DQNs. Monte-Carlo simulations and comparison studies with a state-of-the-art coalition formation algorithm are performed to verify the performance of CA-DQN with varying environmental parameters. The results show that the approach is invariant to the number of detected targets and the number of robots in the swarm. The paper also presents the real-time implementation of CA-DQN for different scenarios using ground robots in a laboratory environment to demonstrate the working of CA-DQN with low-power computing devices.
\end{abstract}

\def\abstractname{Note to Practitioners}
\begin{abstract}
Swarm of robots with an inherent redundancy have potential applications in disaster situations like a forest fire, search, and rescue. While operating in such an environment, it is not easy to establish communication between the robots. However, as a swarm, robots have to stay within each other's vicinity and cooperate. Achieving this without any communication is a challenging problem. In a typical reconnaissance mission, the robots need to search and neutralize an unknown number of heterogeneous targets. It becomes challenging for on-board sensors to give accurate information for various reasons like cluttered space and poor visual visibility. Hence, one needs to develop a decentralized decision control algorithm with information uncertainty to identify the targets and allocate them to robots, followed by neutralization. Also, the scheme should be scalable without much effort. This paper proposes CA-DQN to overcome all the difficulties mentioned above and allow the robots to operate together without any communication to neutralize heterogeneous targets. 
\end{abstract}

\begin{IEEEkeywords}
Reinforcement Learning, Markov's Game, Swarm Robotics, Decision Making, Deep Q-Network (DQN)

\end{IEEEkeywords}

\IEEEpeerreviewmaketitle
\section{Introduction}
\IEEEPARstart
{R}{ecent} advancements in Unmanned Aerial Vehicle (UAV) technologies have accelerated deployment of swarms of UAVs to achieve common goals by coordinating with each other. In the past decade, multiple UAVs have been shown to be effective in surveillance  \cite{FINN2012184},  hazardous operations \cite{Harikumar2019MultiUAVOM, article3},  and in cooperative search missions \cite{Rossi2018ReviewOM,8899680,Harikumar2019MissionAM,article2,article}. Since the decentralized, multiple UAV approaches are equally applicable to ground robots; they are referred to as robots in general. In a reconnaissance mission, swarm-based operations are advantageous due to their inherent decentralized and distributive characteristics. In an unknown/uncertain environment, heterogeneous targets can be neutralized by either a single robot or multiple robots (in a sequence of operations or simultaneous operations). For example, in a typical military reconnaissance mission, targets like transport vehicles can be neutralized by a single robot (referred to as a Single Robot Task -- SRT). However, targets like bunkers require multiple robots to attack (referred to as a Multiple Robots Task -- MRT). Sometimes, multiple robots may be needed to neutralize a target with some time delay between their operations, i.e., in a sequential manner. For example, in disastrous scenarios like firefighting, the robots may need to apply different materials in a particular sequence based on the type of fire. In such missions, communication among robots plays a vital role in cooperation. However, in an uncertain environment, communication may not be available. Thus, the complexity of a multi-robot search problem increases by several folds. Such a mission requires multiple robots to carry out the following operations: a) search and detect an unknown number of targets; b) allocation of targets to the robots; and c) neutralization of the targets by cooperation while having information uncertainty. 

The formulation for a simultaneous search by multi-UAVs and task assignment problem as a single optimization problem has been presented in \cite{Turra,7058448,8393468,Sujit2005MultiUAVTA}. However, in the absence of communication, the articles focus only on tasks involving  SRTs. Once a target is assigned, it is assumed that a robot can generate feasible trajectories and avoid collisions.  Negotiation and consensus-based approaches for the same problem have been proposed in \cite{Gerkey2002SoldAM}, \cite{8628943}. Works like \cite{Meng2008MultiRobotSU, Kanakia2016ModelingMT} use game-theoretic approaches while formulating the problem as a task allocation problem. Note that the above methods are suitable only for SRT applications. In a real environment, one may have heterogeneous targets consisting of both SRTs and MRTs, which may require multi-robot coordination. Further, information uncertainty and partial observability limitations increase the mission complexity.

Recently, multi-robot coordination using a game-theoretic approach is presented as a task allocation problem with information uncertainty in \cite{Dai2019TaskAW}. It uses measurements about the targets to compensate for the intermittent loss of communications. An information-theoretic approach for learning the individual agent policy that mimics the centralized static optimization problem is presented in \cite{Dobbe2017FullyDP}. The above approaches assume a certain level of communication between the agents and are also computationally intensive. It limits their scalability and thus is unable to provide feasible solutions with an increasing number of agents. Recent studies in swarm robotics have focused on applying Deep Reinforcement Learning (DRL) strategies for multi-agent cooperation. Reinforcement Learning techniques, like Q-Learning \cite{watkinsc.j.c.h.dayanp.1992}, define the tasks indirectly using cost functions, which are easy to formulate compared to the modeling of the task \cite{Mataric1997ReinforcementLI}.  These methods make it easier to implement \cite{Yang2004MultiagentRL,Busoniu2010MultiagentRL}; however, the approaches are not scalable and cannot be decentralized as they often require the states of the entire system rather than a single agent. The solution also requires significant memory as the Q-Learning technique specifies different Q-maps \cite{watkinsc.j.c.h.dayanp.1992} for every given state. However, one can reduce the memory requirements by using Deep Q Networks (DQNs) to approximate the optimal policy. Neural network-based solutions have worked well for a single agent problem; however, Multi-Agent Reinforcement Learning (MARL) remains challenging.

The use of deep reinforcement learning (DRL) approaches for cooperative task execution (single target handled by a single agent) seems promising as implemented in \cite{Dibangoye2013OptimallySD, Claes2017DecentralisedOP, Nguyen2017PolicyGW}. Due to the fixed input dimension to the neural network, the solution is not scalable for a varying number of agents and observations over time. Recently in \cite{Liu2017LearningFM, Amato2018DecisionMakingUU, Xiao2019LearningMD}, the issue of scalability is addressed by using a decentralized, partially observable Markov decision process by fixing the information collection region for an agent. However, for better convergence of the network, one needs to generate many training instances in a simulator.  Similarly, \cite{Httenrauch2019DeepRL} and \cite{8903067} address the scalability and convergence issues by developing state representations and reward functions. The above methods are also not applicable to the varying number of agents, as the robots would require retraining. More information regarding decentralized MARL algorithms is given in  \cite{Zhang2019DecentralizedMR}. Deep reinforcement learning has also been used for multi-agent pathfinding \cite{8661608,Ma_2017}, where the agent's final destination is known apriori. It may not be suitable for a search and reconnaissance mission with heterogeneous tasks.

The aforementioned approaches address either target detection problem in an unknown/uncertain environment or task allocation in a territory consisting of homogeneous targets with or without communication. In a typical search mission with heterogeneous targets without communication, there is no integrated solution that addresses the search, task allocation, and motion control for effective collation for neutralization and collision avoidance. Hence, there is still a need to develop an end-to-end solution for a swarm of robots to perform reconnaissance missions in an unknown and uncertain environment.

In this paper, we present a swarm robot solution to search and neutralize heterogeneous targets without communications in an unknown/uncertain environment. A novel algorithm referred to as Context-Aware Deep-Q-Network (CA-DQN), which acts as a decision-making framework for a robot in the swarm, is presented. The environment has an unknown number of heterogeneous targets consisting of single-robot-targets (SRTs) and multi-robot-targets (MRTs). The SRTs require only one robot to neutralize it, while the MRTs require multiple robots to interact with the target in a sequence (one after the other) for neutralizing them. Every individual robot is equipped with a global sensor to detect the targets (their locations and types) and a local sensor to detect their neighboring robots. Depending on both the global and local information, each robot generates a context-aware grid that continuously adapts itself to the changing target distributions. The task allocation algorithm assigns the target and its sequence without communication between the robots. A scenario identification algorithm determines each robot's status as either a 'conflict-free' or a 'conflict' scenario in the swarm, based on the sensor information. At any given time, this context information act as inputs to the Deep-Q-Networks (DQNs) to generate the motion control commands for the robots to detect and neutralize the heterogeneous targets cooperatively. Two different DQNs, (namely, one DQN for a conflict scenario and another DQN for a conflict-free scenario) are used for the action-value approximations needed in the reinforcement learning context. This context-awareness scheme enables a robot to use local and global information to identify the corresponding scenarios and take appropriate actions. 

Further, a detailed performance evaluation of the proposed CA-DQN using Monte Carlo simulations with varying target distributions and swarm size is presented. Also, a comparison with the coalition formation approach [34] that uses sub-team formation [35] for decentralized neutralization of a target in a simultaneous manner is presented. However, for comparison purposes, [34] has been modified for sequential task execution, allowing MRT neutralization. The study results show the efficiency of CA-DQN for a swarm to perform search and neutralize missions under information uncertainty and no communication. Also, experimental studies in the laboratory conditions are performed to evaluate the performance of CA-DQN upon deployment in real-time. The experimental study results show that CA-DQN can be implemented with low-power computing devices, which form the fundamental basis of swarm robotics. The simulations and real-world experiments show that CA-DQN is a scalable decentralized approach. It is invariant to the number of targets detected and the number of robots in the swarm. Finally, CA-DQN is an end-to-end solution for a search and neutralization mission. 

The paper is organized as follows: Section \ref{sec:2} presents the mathematical formulation followed by the proposed decision-making framework. Section \ref{sec:3} discusses the performance of the proposed CA-DQN using Monte Carlo simulations and a comparison with the modified state-of-the-art coalition formation based method. Section \ref{sec:4} provides the performance results for the CA-DQN using an experimental setup. Finally, the conclusions from the studies are summarised in section \ref{sec:5}.

\section{Decentralized cooperative swarm for search and neutralization missions}\label{sec:2}
In this section, the problem definition of the search and neutralization mission is provided, followed by the problem setup. Next, the algorithm for information embedding in context regions and scenario identification is given. Finally, the details regarding the deep-Q-network architectures and their training procedure for multi-robot coordination is provided.

\subsection{Problem definition}
\begin{figure}
	\centering
    \includegraphics[width=0.9\linewidth]{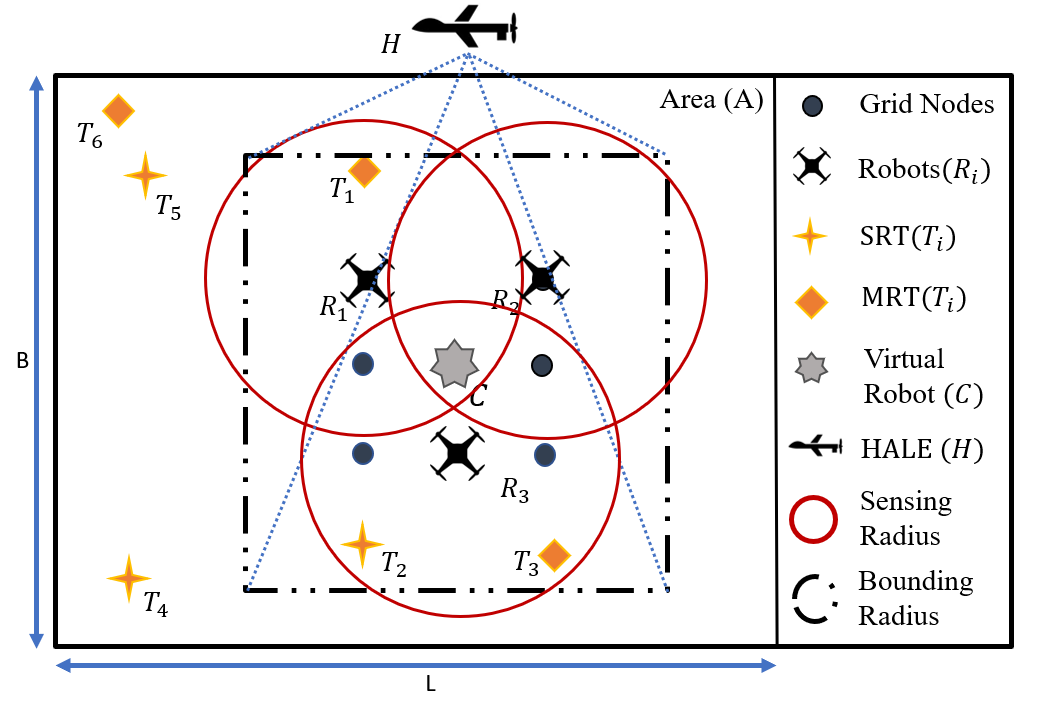}
    \caption{A typical scenario of a swarm based search in unknown/uncertain environment.}
    \label{fig:scenario}
\end{figure}

A typical search and neutralization mission for heterogeneous targets (consisting of single and multi-robot targets)  by a swarm of robots in a given unknown/uncertain environment, as shown in fig. \ref{fig:scenario} is considered for the study. The mathematical formulation of the problem is first presented, followed by the proposed decentralized decision control framework for solving the problem.

The mission is carried out in a bounded region $\mathcal{A}$ using $N$ robots. The region contains $M$ targets whose initial positions are unknown. Each robot is represented as  $R_i,i=1,\ldots,N$. The robots can carry $n$ different types of resources in limited quantities for neutralising the targets. The
robot capability vector $\mathcal{J}_i$ is of the form,
\begin{eqnarray}
\mathcal{J}_i = <J_{i}^1,\ldots,J_{i}^n>,~~~ i =1,\ldots,N
\end{eqnarray}
where $J_i^p,~p=1,\ldots,n$ represents the number of $p$ type resources held by the robot $R_i$. The robots are operating as a swam under the following constraints.
\begin{itemize}
	\item[a)] At any given time, multiple robots should not occupy the same space, and
	\item [b)] There is no communication network available for the robots to share their information. 
\end{itemize}

The robots are equipped with two types of sensors, a) global sensors with a range of $r_g$ and b) local sensors with a range of $r_l$ to detect their neighbors, considering that $r_g >> r_l$. Without loss of generality, it is assumed that the robot motion and terrain do not affect the sensor ranges. When $R_i$ detects a target $T_j$, it is assumed that it can ascertain the type of resource required to neutralize it. If $m$ different types of resources and quantities are required to engage the target $T_j$, then the target resource requirement vector is represented as,
\begin{eqnarray}
\mathcal{L}_j = <L_{j}^1,\ldots,L_{j}^m>,~~~ j =1,\ldots,M
\end{eqnarray}
where $L_{j}^q$ represents the quantity of type-$q$ resources required to neutralize the target $T_j$, and $M$ is the total number of detected targets at time $t$. When $L_j^q=1$, then the target is denoted as SRT; otherwise, it is denoted as an MRT. The MRTs require multiple robots to operate in a sequence to neutralize them. The task assignment provides inputs to the robots on the required sequence of neutralization, and the individual robots have a memory to avoid neutralizing the same target again. The robots use the global sensor to identify the state of a target (i.e., whether the target is live or neutralized or the current state of an MRT after being neutralized by any robot).

\begin{figure}
	\centering
    \includegraphics[width=0.6\linewidth]{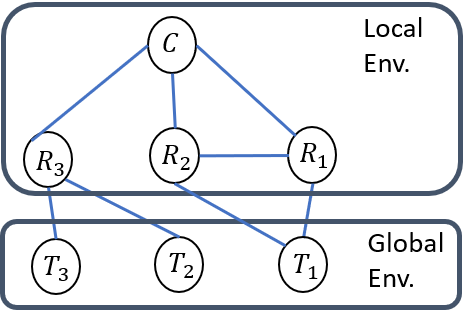}
    \caption{Global/Local view of all the robots for the scenario in Fig. \ref{fig:scenario}}
    \label{fig:defination_graph}
\end{figure}

The robots are subjected to kinematic constraints and move at a speed of $v_i$. The kinematic equations  of the robots are modeled as, 
\begin{equation}
\dot x_i = v_i\cos \psi_i; \ 
\dot y_i = v_i\sin \psi_i; \ 
\dot \psi_i = k(\psi_i^d - \psi_i)
\end{equation}

where $\psi_i^d$ is the desired heading of the robot,  $\psi_i$ is the current heading (the angle the robot velocity vector makes with the $x$-axis), and $k$ is the gain. The  heading rate is constrained to $-\omega_{\max} \leq \dot\psi \leq \omega_{\max}$.

The mission needs to be accomplished in minimum time.  Let $\mathcal{T}$ represents the minimum mission time that the robots take to neutralize all the targets. In the given environment, the target locations and their resources are not known \textit{apriori}. Therefore, the robots have to carry out a search mission to detect the targets and neutralize them. Let us assume that the robots use a strategy $\mathbb S$ to neutralize an MRT. Then, the total time taken to neutralize all the targets is given by,
\begin{eqnarray}\label{eqn:missionTime}
\mathcal{T}(\mathbb S)\leftarrow \mathcal{T}_s + \sum_{j=1}^M \mathcal{T}_j^{\mathbb S},
\end{eqnarray}  
where $ \mathcal{T}(\mathbb S) $ represents the mission time when the robots use the strategy $\mathbb S$ to neutralize an MRT, $\mathcal{T}_s$ represents the time consumed by the search scheme to detect the targets, and $ \mathcal{T}_j^\mathbb{S}  $ represents the time taken to neutralize the target $T_j$ when using  $\mathbb S$ strategy.  As the environment knowledge is unknown and uncertain, the best alternative mechanism is to determine a strategy $\mathbb S$ for a given search scheme such that the mission time  $\mathcal{T}(\mathbb S)$ is minimized.

Fig. \ref{fig:scenario} shows a typical scenario of a search environment with a swarm of robots. The $N$ robots are deployed to search and neutralize the $M$ heterogeneous targets. For ease of understanding, the range of both the global and local sensor is shown to be the same, against the assumption that $r_g >> r_l$. The environment consists of both SRTs and MRTs. It is assumed that a High-Altitude Surveillance Vehicle (HALE) monitors the swarm of robots. The HALE identifies each robot's position and orientation to navigate itself to the centroid of the swarm. It may be noted that the HALE is also a robot whose existence is considered in the swarm as a virtual robot, as shown in fig. \ref{fig:scenario}. For the swarm to stay intact, each robot needs to stay within a fixed radius from the position of the virtual robot ($C$). To determine the virtual robot's position, it is also assumed that every robot is equipped with an upward-looking sensor to find HALE's position. This positional information helps the robots in restricting their motions within the bounding region of the swarm. It may be noted that individual robots decide on their search direction, select tasks, and coordinate with other robots to neutralize heterogeneous targets based on only the sensor information in a decentralized manner.

Fig. \ref{fig:defination_graph} presents both the global and local view of individual robots (a robot's visibility of the targets and other neighboring robots) of the swarm for the scenario depicted in fig. \ref{fig:scenario}. In fig. \ref{fig:scenario}, $R_1$ is able to detect neighbour $R_2$ and also detect a MRT target $T_1$, whereas $R_3$ detects two targets ($T_2$ and $T_3$). Similarly, $R_2$ can see its neighbour $R_1$ and MRT $T_1$. All robots can detect the virtual robot ($C$), which represents HALE within the swarm. The figure clearly illustrates the information uncertainties between the robots and targets present in the scenario. It should be noted that at any given time, the information about the target and the neighboring robots within the sensing range is used to define the robot's state within the swarm. The information (both global and local) structure is dynamic, and hence, the decision-making ability of robots should be adaptive and scalable.
In summary, the problem considered in this paper is to detect the maximum number of targets, allocate different tasks to individual robots, and neutralize these targets by cooperating/coordinating with other robots within the swarm in minimum time.

\subsection{Context-Aware Deep Q-Network (CA-DQN)}
This section presents a decision-making framework for a robot to perform the required mission. The decision-framework uses the context information and a Deep-Q-Networks to provide the necessary decisions to individual robots to complete the overall mission in minimum $\mathcal{T}(S)$. The proposed framework is referred to as the Context-Aware Deep Q-Network (CA-DQN).

\begin{figure}
	\centering
    \includegraphics[width=1\linewidth]{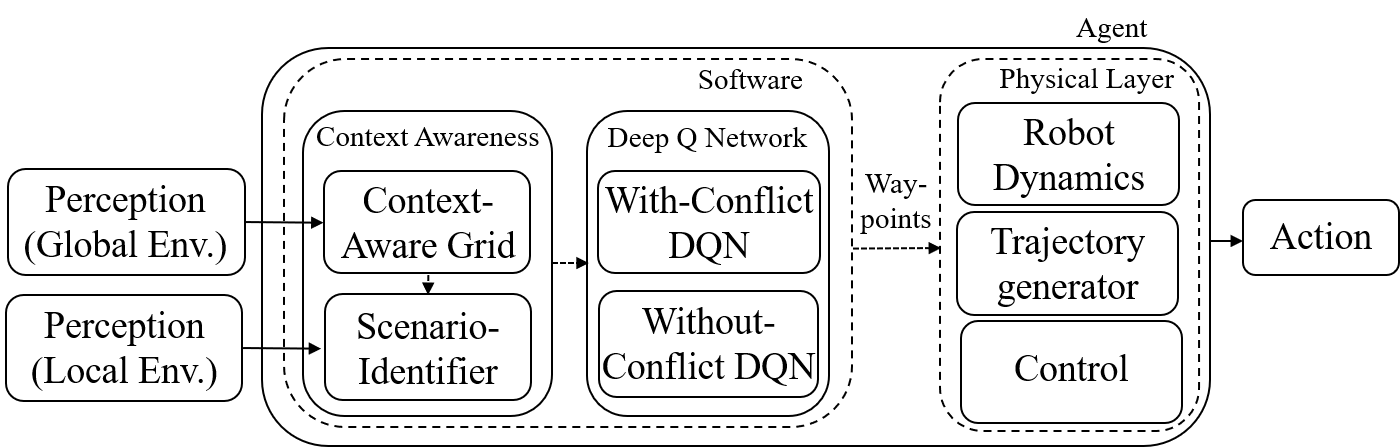}
    \caption{Schematic diagram of CA-DQN.}
    \label{fig:schematic}
\end{figure}

Fig. \ref{fig:schematic} shows the schematic of the proposed CA-DQN and its information flow. Each robot senses the global environment to detect the target distribution and the local environment to detect its neighboring robots. While moving as a swarm, once a target is identified, the robots obtain information about the position and type of target. This information is then used for allocating the targets to the robots based on the minimum distance between them. Therefore, the task allocation ensures that an MRT is neutralized using a sequence that is determined based on distances. At any given time, the number of targets detected by the robot $R_k$ is $m_k$, and the number of detected neighbors is $n_k$. The robot $R_k$ generates a cost matrix $(C_k)$ based on the distances between the detected robots and the number of detected targets at any given time. The robots carry out task allocation by solving the following:
\begin{equation}
    \label{eq:task}
    minimize \sum_{k \in n_k} \sum_{j \in m_k} C_{kj} x_{kj}  
\end{equation}
\begin{equation}
    \label{eq:task_subject}
    subject \hspace{0.8ex} to  \sum_{j \in n_k} x_{kj} \leqslant 1 ;  \sum_{k \in m_k} x_{kj} \leqslant L_j^q  
\end{equation}
where $L_j^q$ represents the quantity of type $q$ resources required to neutralize the target $T_j$. The above optimization problem provides task allocation to the robots in the swarm in a decentralized manner.

After the targets are allocated, depending upon the target location and type, the swarm needs to get in the required formation to neutralize the targets. As multiple robots are operating in the environment and share the same global coordinate system for localization, the actions taken by individual robots may be conflicting with each other, as the robots cannot occupy the same space at any given time. Therefore, to maneuver in a collision-free manner within the swarm, each robot generates its own grid, called a context-aware grid, by keeping the HALE at its centroid. The grid nodes represent points in the vicinity of the robot in a global coordinate system; hence the positioning of these nodes is local to every individual robot. At any given time, $t$, every robot and targets are assigned a unique node in its context-aware grid.  The context information is embedded into the grid, which is then used by the robots to identify the scenario in which they are present. Based on the scenario, two different DQN's (for 'conflict' and 'conflict-free' scenarios) are trained to ensure collision-free navigation to the target location. The output of DQN is a probabilistic array that maps the utility of a discrete set of actions in a given scenario. These values are then used to change the robot's node assignment to the next adjacent grid node to maneuver in its grid.  As every node of the context-aware grid is associated with a global coordinate, these locations act as waypoints for the robots to follow. The robot's motion is controlled using a proportional-integral (PI) \cite{10.5555/516039} controller for achieving the desired objectives while keeping the swarm intact. Although the localization is global, the grid nodes generated are local to every individual robot. It helps the robot to obtain both local and global information for operating in the environment. Next, the individual components of the CA-DQN are described.

\subsubsection{Context awareness scheme}

In the current problem setup, there are $N$ homogeneous robots in a swarm performing search and neutralize mission in an unknown and uncertain environment. Every robot generates its grid by keeping the virtual robot at the centroid; and assigns unique nodes for itself, targets, and the neighboring robots. Grid nodes are not generated beyond the radius of the swarm boundary and the search area boundary. As the environment is dynamic, the target distribution may not coincide with a uniform grid (i.e., a grid with a constant distance $d$ between the grid nodes). 
This work assumes that a robot needs to be over the target to be able to neutralize it. A robot also uses the grid nodes as waypoints to be over a target and avoid neighboring robots. Therefore, the context-aware grid is deformed according to the target distribution and neighboring robots to ensure that the grid nodes overlap with the object's position. The distances between the grid nodes are stored and modified to deform and reconstruct the grid. It may be noted that if multiple objects are equidistant from a given node, the grid is modified to occupy any one of the nearest objects using their identity as the priority order.

\begin{figure}
	\centering
    \includegraphics[width=0.7\linewidth]{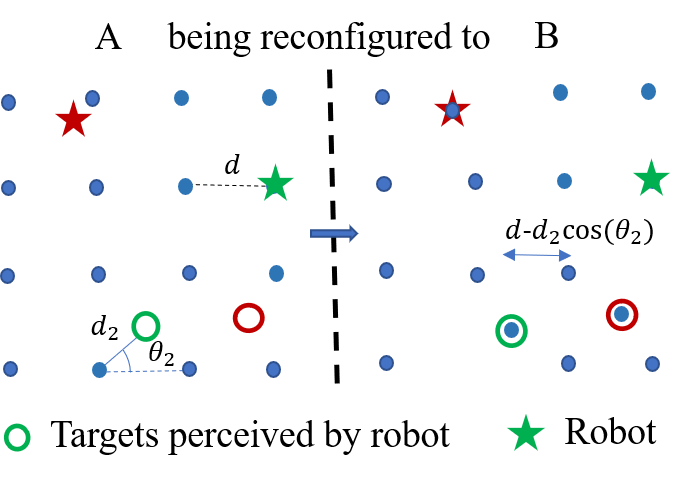}
    \caption{Context-aware grid: Grid adaptation from configuration A to B based on detected targets and neighbouring robots.}
    \label{fig:Dmat_math}
\end{figure}

Two matrices $d_x$ and $d_y$ are defined to store the corresponding distances between the nodes along the column and rows (i.e., the x and y-axis, respectively, in the robot's frame of reference). To modify the grid, based upon the context information, the distance elements in these matrices are updated as $d_x^{new} = d \pm d' \cos(\theta)$ and $d_y^{new} = d \pm d' \sin(\theta)$. Here, $\theta$ represents the angles made by the object location with the nearest node location. The term $d'$ represents the distance between them. Fig. \ref{fig:Dmat_math} shows a typical context-aware grid of a robot ($R2$) before and after deforming from configuration $A$ to $B$. The blue dots represent the context grid nodes. Consider, the distances between the sensed objects and their nearest grid nodes to be $d_1$, $d_2$ and $d_3$ with the angle between them being $\theta_1$,$\theta_2$ and $\theta_3$ respectively. For simplicity, only one distance ($d_2$) and angle($\theta_2$) is shown in the figure, considering the rest of the distances similarly. The $d_x$ and $d_y$ matrices for the deformed grid $B$ are given by,

\begin{align}
\label{eq:25}
d_x &=
\begin{bmatrix}
d - d_3 \cos(\theta_3) & d + d_3 \cos(\theta_3) & d\\
d & d & d\\
d & d & d - d_1 \cos(\theta_1)\\
d + d_2 \cos(\theta_2) & d - d_2 \cos(\theta_2) & d
\end{bmatrix}\\
d_y &=
\begin{bmatrix}
d & d - d_3 \sin(\theta_3) & d & d\\
d & d & d & d + d_1 \sin(\theta_1) \\
d & d - d_2 \sin(\theta_2)  & d & d - d_1 \sin(\theta_1) 
\end{bmatrix}.
\end{align}

A robot can sense the HALE to estimate the HALE's global coordinates and obtain its relative position. Using the HALE's coordinates as the centroid of the grid, the robot uses the $d_x$ and $d_y$ matrices to obtain the global coordinate of the grid nodes assigned to it. These coordinates then act as the waypoints for the robot to follow.
It may be noted that the algorithm to detect the position of the target has a Gaussian uncertainty. It will introduce uncertainty in the position of the grid nodes when the targets are far away. However, uncertainty reduces significantly when the targets are closer to the robot. Hence, it initially affects the robot trajectory, but it will be negligible when it is closer to the target. 

\subsubsection{Scenario-Identifier algorithm}

As mentioned earlier, the actions taken by individual robots may be conflicting with each other as multiple robots share the same global coordinate system for localization. After the task has been allocated, the Scenario-Identifier (SI) algorithm aims to classify a robot's current situation either as a 'conflict' or a 'conflict-free' scenario for dealing with them separately.

\begin{figure*}
\centering
    \subcaptionbox{Context region towards assigned target.\label{fig:a_r}}
    {\includegraphics[width=0.31\textwidth]{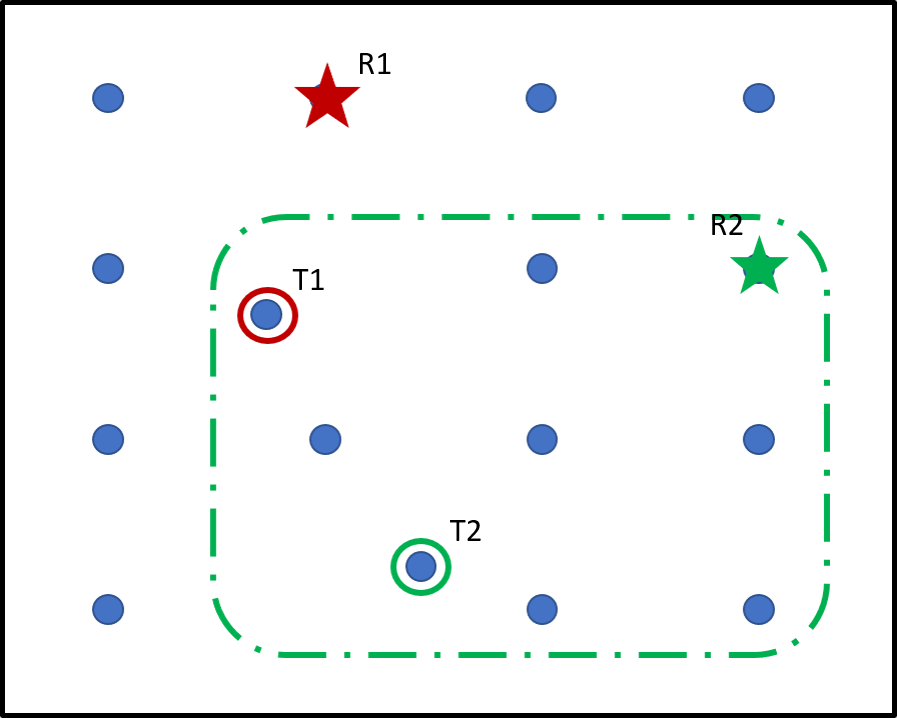}}
    \subcaptionbox{Context region towards other target\label{fig:b_r}}
    {\includegraphics[width=0.31\textwidth]{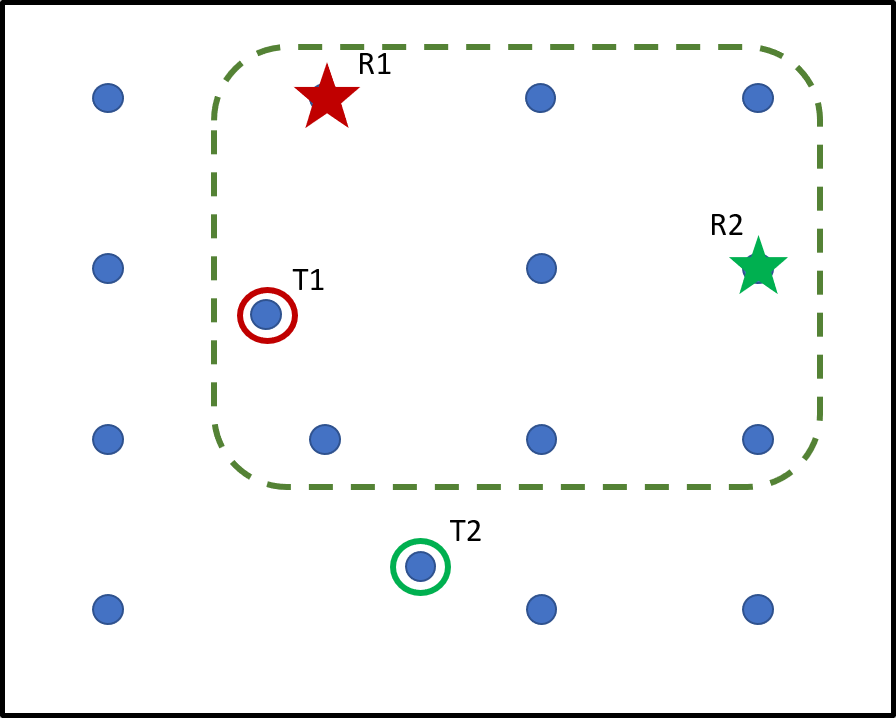}}
    \subcaptionbox{$2\times2$ region to check for conflicts.\label{fig:c_r}}
    {\includegraphics[width=0.31\textwidth]{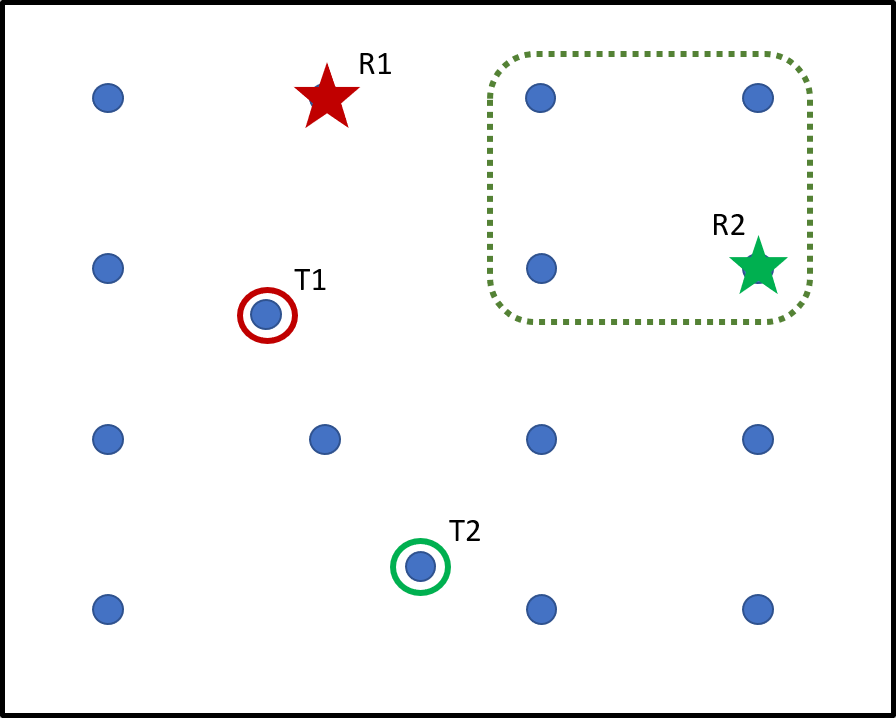}}
    
    \vspace{+1.5ex}
    \caption{A depiction of targets and neighbouring robots information embedded in a adaptive grid representation for SI algorithm}
    \label{fig:ROI}
\end{figure*}

Every robot has its context-aware grid that adapts to any given distribution of detected targets and neighbors. The scenario identifier (SI) algorithm uses the information embedded in smaller context regions in the context-aware grid to determine the robot's situation. For the scenario in fig. \ref{fig:ROI}, let us assume the robot $R_1$ is assigned to neutralize the SRT $T_1$ and the robot $R_2$ is assigned to neutralize the SRT $T_2$. The figure highlights different types of context regions generated by the SI algorithm for the robot $R_2$. First, the SI algorithm generates a $3\times3$ context region towards its assigned target, as shown in fig. \ref{fig:a_r}. In this region, the robot checks for the presence of other targets or robots. If no other robots and target other than the assigned one exist in this region, the robot classifies this scenario as 'conflict-free'. Hence, it can proceed to move toward the target without bothering about the rest of the surroundings. From fig. \ref{fig:a_r}, it can be observed that target $T_1$ is present in the context region of $R_2$. Hence, the robot generates another $3 \times 3$ context region in the context-aware grid towards target $T_1$, as shown in fig. \ref{fig:b_r}. If there is no robot found in this region, the robot masks the target (i.e., considers it as an obstacle) and proceeds further. Since the robot $R_1$ is also present in the context region in fig. \ref{fig:c_r}, $R_2$ generates a $2 \times 2$ context region towards the robot. If the $R_1$ is found in this $2 \times 2$ region of $R_2$, the scenario is classified as a 'conflict' scenario for $R_2$. As that is not the case in \ref{fig:c_r} the scenario is classified as 'conflict-free' for $R_2$. The grid nodes assigned to $R_1$ and $T_1$ are masked and considered an obstacle. It would effectively make $R_2$ to choose a motion control command that does not lead to the grid nodes of the obstacles.
The explanation of the SI algorithm is summarized and given in Algorithm \ref{Algo2}.

\begin{algorithm}

\caption{Pseudocode of SI Algorithm}
\begin{algorithmic}

\Function{conflict with robot}{} \funclabel{alg:a} \label{alg:a-line}
    \State Generate new $2\times2$ Context region on the grid;
    \If{presence of other robots}
        \State assign as 'conflict scenario';
        
    \State assign $2\times2$ context region;

 \Else
    \State mask the robot;
    \State assign as a conflict-free scenario;
    \State use original $2\times2$ context region; 

    \EndIf
\EndFunction

\Statex
 \State \textbf{Task Allocation}: using eq. \ref{eq:task} \& eq. \ref{eq:task_subject};
 \State \textbf{Input}: Generate $3\times3$ context region;
 \State Scenario Identifier procedure:
 \If{presence of any other targets}
    \State generate new $3\times3$ context region for each target;
    \If {presence of any robots}
        \State call \Call{conflict with robot}{};
    \Else 
        \State mask target location;
        \State assign as a conflict-free scenario
    \EndIf
 \EndIf
 \If {presence of any robots}
    \State call \Call{conflict with robot}{};
 \Else
    \State assign as a conflict-free scenario;
 \EndIf 
 
\end{algorithmic}
\label{Algo2}
\end{algorithm}

\subsection{Deep Q-Networks and their learning algorithm}
\label{sec:learning}

In this work, the communication free coordination within the swarm has been formulated as a Markov's game \cite{richards.suttonandrewg.barto1998}. At every time step, the robot perceives its state $s$ to take action $a$ using a policy $\pi(a|s)$ and thus fetch a reward $r$ for the transition to a new state $s'$. In a Markov's game, this is represented as an experience tuple given by $\langle{s,a,r,s'}\rangle$. The robot must learn the optimal policy $\pi^*(a|s)$ for any given state to obtain the desired action. To obtain the optimal policy $\pi^*(a|s)$, the Q-Learning technique developed in \cite{watkinsc.j.c.h.dayanp.1992}, uses Q-value approximations. The action-value function $Q(s,a)$ determines the utility of a probable $a$ when a robot is in $s$. In this paper, a deep reinforcement learning framework is used where a Deep Neural Network (DQN) is used to approximate the $Q$-value function for the probable state-action pairs, rather than analytically developing it \cite{article5}. The robot chooses the action corresponding to the maximum $Q$-value for its motion control command. In every $i^{th}$ iteration, an experience tuple $\langle{s, a,r,s'}\rangle$ is sampled. The training process involves two different DQNs with parameters $\theta$ and $\theta'$. One is used to obtain the $Q$ values for the actions, while the other includes all the training updates and is called the target network. After every $k$ iterations, $\theta'$ and $\theta$ are synchronized. The network parameters $\theta$ is updated in every iteration using the backpropagation algorithm to minimize the loss function given by
\begin{equation}
    L_i(\theta_i) = (r + \gamma \max Q(s',a|\theta^{'}) - Q(s,a|\theta))^2
    \label{eqn:loss}
\end{equation}
where, $\gamma$ is the discount factor used for the training process, and $r$ is the reward. Usually, the DQN uses a $\epsilon$ greedy strategy during the training process due to its empirical success. Studies in the literature by \cite{Mnih2013PlayingAW, Bellemare2013TheAL} have highlighted the theoretical convergence of the DQN in a geometrical rate \cite{Yang2019ATA}. Hence, in this paper, the $\epsilon$ greedy strategy has been adopted for training the DQN. Next, the state representations and the DQN architectures for both the 'conflict' and 'conflict-free' scenarios are described.

\subsubsection{Deep Q-Network for a conflict scenario}
\label{subsec:1}
\begin{figure}
	\centering
    \includegraphics[width=0.8\linewidth]{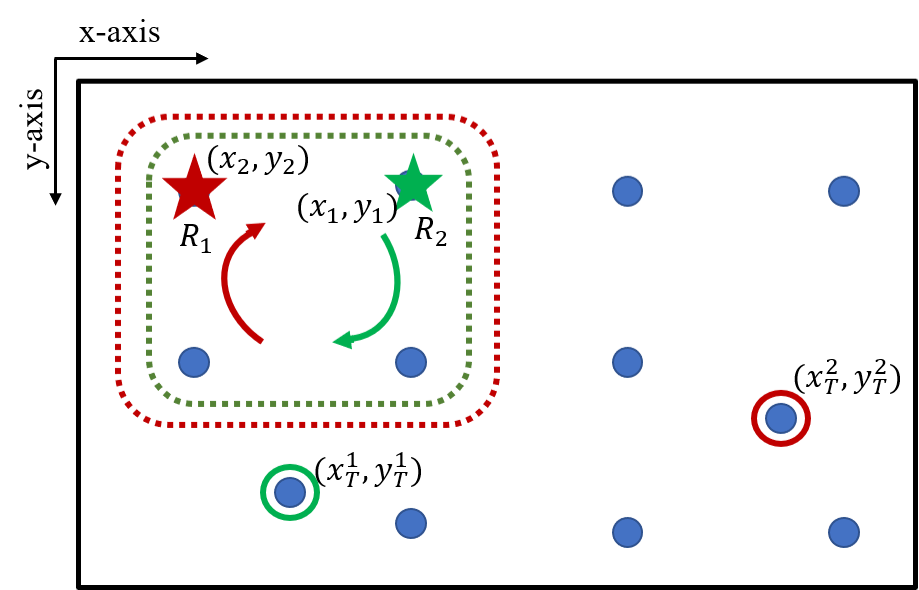}
    \caption{A conflict scenario between two robots and their corresponding $2\times2$ context regions.}
    \label{fig:state}
\end{figure}

In this subsection, the process of extracting the state information from the context region is described, followed by the DQN architecture used for the Q-value approximation. For a 'conflict' scenario, the context region is a $2 \times 2$ grid around the robot.  A typical 'conflict scenario' is shown in fig. \ref{fig:state} where the $2\times2$ context region for both the robot $R_1$ and $R_2$ is shown. The information in this context region is used to represent the state of a robot, which then acts as the input to the DQN. In this work, the number of possible actions, $n_a = 5$ is considered, i.e., $\left\{a_l,a_u,a_r,a_b,a_s\right\} \in \Re^5$. However, depending upon the scenario and the context information, the robot's available action space can be lower than or equal to $5$. For example, in fig. \ref{fig:state}, the available action for $R_1$ is $\left\{a_r,a_b,a_s\right\}$, and similarly for $R_2$ is  $\left\{a_l,a_b,a_s\right\}$. The state of the robot ($\bf{s}$) is represented by traversing in a clockwise direction in the $2 \times 2$ context region. Each node in the grid is represented by a $3 \times 1$ vector. If the node is occupied by the robot itself then the information is coded as $\left [ \lfloor (x^T-x)/|x^T-x| \rfloor,  \lfloor (y^T-y)/|y^T-y| \rfloor, 1 \right ]$, where ($x$,$y$) is the robot position and ($x^T$,$y^T$) is its target position. It may be noted that these coordinates represent a node's position in the grid, with the top-left node being the origin. Similarly, if the node is occupied by other robots then information is coded as $\left [ \lfloor (x_n^T-x_n)/|x_n^T-x_n| \rfloor,  \lfloor (y_n^T-y_n)/|y_n^T-y_n| \rfloor, 0 \right ]$, where ($x_n$,$y_n$) is the other robot position and ($x_n^T$,$y_n^T$) is its target position. The empty nodes in the context grid are represented using a null vector $[0,0,0]$. A typical 'conflict scenario' is shown in fig. \ref{fig:state} where the $2\times2$ context region for both the robot $R_1$ and $R_2$ robot matches. The robots and their targets are marked in the figure. The context region in the grid has got two blank nodes as there are no objects nearby. Considering the axes as shown in the figure, using the information about each node within this region, the state vector for $R_1$ can be given as $\it{s}_1=[1,-1,1,-1,1,0,0,0,0,0,0,0]$ and for $R_2$, $\it{s}_2=[-1,1,1,0,0,0,0,0,0,1,-1,0]$.

\begin{figure}
	\centering
    \includegraphics[width=1\linewidth]{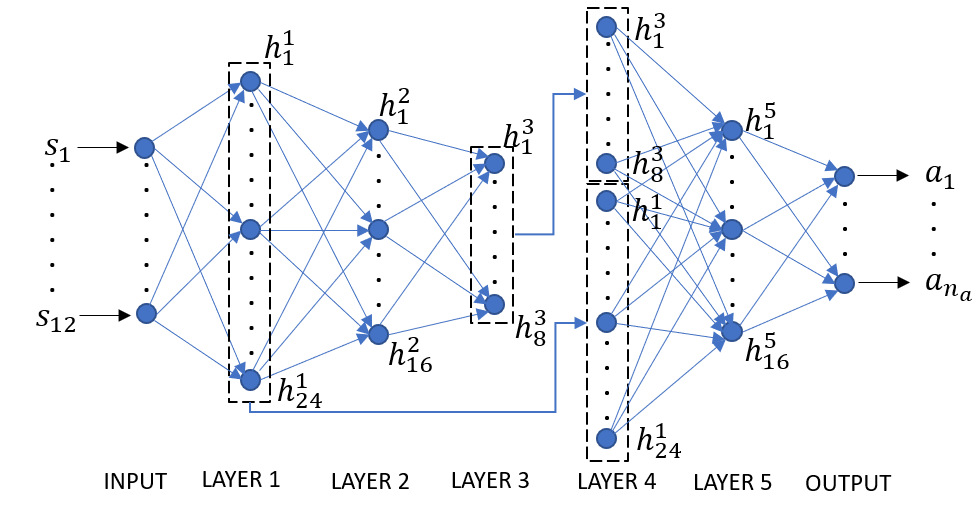}
    \caption{Architecture for DQN in a conflicting scenario.}
    \label{fig:Coop_arch}
\end{figure}

The architecture of the DQN for a conflict scenario is shown in fig. \ref{fig:Coop_arch}. The DQN approximates the Q-function between the state ($\bf{s} \in \Re^{12}$) and the action state ($\bf{a} \in \Re^{n_a}$) where, $n_a$ is the number of action states. The DQN has five hidden layers, and layer $4$ receives the outputs from layer $3$ and layer $1$. The neurons in the hidden layer (except layer $4$) employ hyperbolic tangent function and approximate the nonlinear Q-function relationship.  The outputs of neurons in the hidden layers are given as,
\begin{equation}
    h^l_i = \tanh \left(\sum_{j=1}^{N^{l-1}} w^l_{ij} h^{l-1}_j\right)
\end{equation}
where $i=1,2,\cdots,N^l$ ($N^l$ is the number of neurons in the $l$th layer) and $l = 1,2,3,5$. At $l= 4$, the architecture has an augmented layer which is a concatenation of layer $3$ and $1$ of the network, i.e., $h^4 = \left[h^3 h^1\right]$. This is used to ensure a proper flow of gradient in the network during back propagation \cite{kelley1960gradient}. At $l=0$, the neuron output is the state of the robot, i.e., $h^0_j = s_j$. The $W^l$ represents the weight connection between $l$ and $l-1 ^{th}$  layer. A Softmax activation function is used over the output layer to obtain a probabilistic output.

For a given state-action pair ($s,a$), predicted by the DQN, a reward is generated, and the weights of the network are updated as per the training process mentioned in section \ref{sec:learning}. 
For the 'conflict' scenario, one can have a maximum of four robots in the local context region as it is of size $2 \times 2$. For training, the self play-in approach is adopted, where individual agents' rewards are used for training the same instance of the DQN (policy). However, during deployment (i.e., after training), the robots use different instances of the DQN of their own. Therefore, the robots take actions on their own, without any information communicated among each other.

\begin{figure}
	\centering
    \includegraphics[clip,trim=0mm 0mm 5mm 5mm,,width=0.9\linewidth]{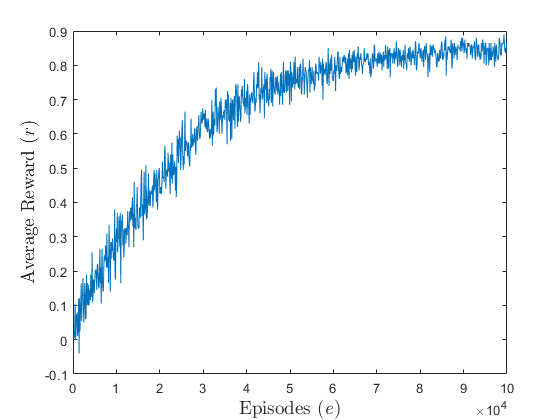}
    \caption{Average reward value per 100 episodes of DQN for conflict scenario.}
    \label{fig:reward_conv}
\end{figure}

In the training process, two robots are made to play in the context region. A positive reward of $+1$ is provided for every iteration passed without collision and an additional $-0.1$ to encourage completion in minimum time. A negative reward of $-1$ is awarded for a collision. As mentioned above, a maximum of $4$ robots can participate in such a scenario. Therefore the training process is repeated for three and four robots by considering the trained policy of their previous cases, respectively. To study the empirical convergence of the  DQN algorithm, the average reward of over $100$ games is computed. The average reward value in every $100$ iteration is shown in fig. \ref{fig:reward_conv}.
The maximum reward one can achieve for a conflict-scenario is $0.9$. As shown in the figure, the average reward saturates close to $0.9$. Also, it may be noted that the average reward is noisy as the small changes to the weights of policy can lead to large changes in the policy's state distributions. The convergence study indicates that stable training of DQN can be achieved using reinforcement learning methods.

\subsubsection{Deep Q-Network for conflict-free scenario}

For the conflict-free scenario, a context region of a $3\times3$ around the robot within the context-aware grid is considered. Similar to that of the 'conflict' scenario, the information within this region is used to represent the state ($\bf{s} \in \Re^2$) of the robot. The state is represented as a $2\times1$ vector, with the information coded as $\left [ \lfloor (x^T-x)/|x^T-x| \rfloor,  \lfloor (y^T-y)/|y^T-y| \rfloor \right ]$ and is used as an input to a DQN. Then the DQN is used to approximate the Q value function between the state ($\bf{s} \in \Re^{2}$) and action space ($\bf{a} \in \Re^{n_a}$). The DQN used in this case has two hidden layers with the hyperbolic tangent function as the activation function. The softmax activation function is used in the output layer of dimension $a_n$ to obtain a probabilistic output.
The robot fetches rewards for a given state-action pair. A negative reward of $-0.1$ is provided with every iteration passed without the robot reaching the target. Otherwise,  a positive reward of $+1$ is assigned to the robot.  With these reward values, the loss is calculated for the network by using Eq. \ref{eqn:loss}, and the weights of the network are updated using backpropagation. It may be noted that one can also use a greedy algorithm to find the shortest path \cite{dij} for the robots in conflict-free scenarios. However, to maintain a unified platform for both cases, a DQN has been used here.

\section{Performance Evaluation of CA-DQN}\label{sec:3}

In this section, the working of CA-DQN is first presented, followed by the performance evaluation of CA-DQN using Monte Carlo simulations and real-world experiments. For the simulation study, a swarm of robots performing search and neutralization missions in a bounded search area ($A$) of $200m  \times 200m $ is considered. The robots used in the swarm are assumed to be equipped with global sensors that can detect targets within a $20m$ radius and local sensors to detect neighbors within a $10m$ radius. The targets are randomly distributed in $A$. Their locations and types remain unknown to the robots until the time the targets are detected. The maximum speed of the robot is constrained to $15 m/s$. The simulations are carried out in a Python 2.7 environment on an Intel Core-i7, a 2.6-GHz processor with 16-GB RAM.

\subsection{Working of CA-DQN}

\begin{figure}
	\centering
    \includegraphics[clip,trim=1mm 2mm 2mm 2mm,width=0.85\linewidth]{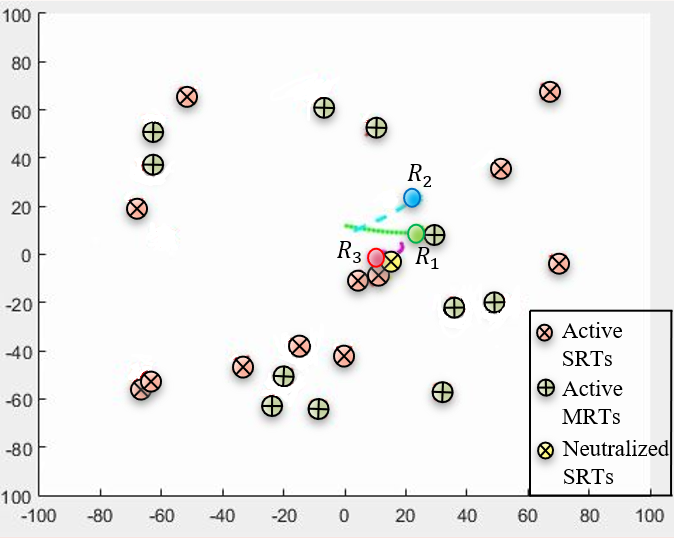}
    \caption{Initial configuration of the robots and targets in a simulation of search and reconnaissance mission.}
    \label{fig:search31}
\end{figure}
\begin{figure}
	\centering
    \includegraphics[clip,trim=1mm 2mm 2mm 2mm,width=0.85\linewidth]{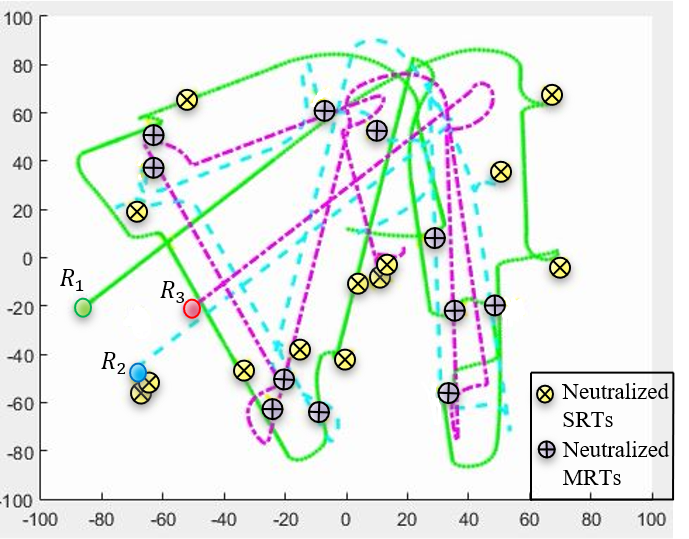}
    \caption{The path traversed by robots depicting their motion in a simulation of search and reconnaissance mission.}
    \label{fig:search32}
\end{figure}

To further understand the working of CA-DQN, consider an example with a swarm of three robots ($R_1$, $R_2$, and $R_3$) performing search and neutralize mission. The environment consists of $24$ heterogeneous targets that are randomly initialized over the arena, as shown in fig. \ref{fig:search31}. The robots are initialized within a $20m \times 20m$ region near the origin $(0,0)$ with random positions and headings. Each robot then generates the context-aware grid in its vicinity with HALE at the centroid. It is assumed that the robot can detect the search area boundary and swarm boundary. Grid points are not generated beyond these boundaries. As the robots search the area, the information about neighboring robots and targets from its local sensors and global sensors are mapped to the context-aware grid by deforming it using $d_x$ and $d_y$ matrices. It ensures that the grid nodes overlap with the detected objects' positions, as the grid nodes represent points in the global coordinate system. Therefore as the swarm moves in the arena, the context-aware grid keeps the swarm intact and helps the robots maneuver within the swarm in a collision-free manner to neutralize the targets. At any given time, a robot has an assigned node and target node. The global coordinates of the assigned node are the waypoints for the robot to maneuver in the global frame, and the target node is used to move within the swarm. When no targets are detected, the robot selects a target node away from the neighboring robots' to maximize the search area coverage.
Upon detecting a target, the context-aware grid node closest to the target is considered the target node. As the robot must move from the assigned node to the target node within the swarm, conflicting situations can occur with neighboring robots (as two robots cannot occupy the same space). Therefore in every time step, the SI algorithm identifies the scenario as either 'conflict-free' or 'conflict' scenario to deal with them separately. The robot uses the DQNs defined for the corresponding scenario to move from the assigned node to the target node. Depending upon the classification, the state representation is generated and acts as an input to the DQN. At any given time, the number of possible actions $n_a = 5$, i.e., $\left\{a_l,a_u,a_r,a_b,a_s\right\} \in \Re^5$ is considered (a robot can move left, up, right, down and stay in a grid). However, as given in section \ref{subsec:1}, depending upon the scenario and the context information, the robot's available action space can be lower than or equal to 5. As the output of DQNs are the Q-values of the actions in a given scenario, the maximum of Q values corresponding to the available action space is taken as the robot's final action. This action selects an immediate neighboring node in the context grid for the robot to maneuver. Since the nodes represent points in the global coordinate system, they form a waypoint for the robot to follow. The PI controller gains are set as $Kp = 0.2$ and $Ki = 0.003$. The angle formed by the line joining the robot's current position and the assigned node, with respect to $x$-axis is the desired heading. The mission terminates when all the targets in the given area are neutralized. The individual robots' trajectories are highlighted in fig. \ref{fig:search32} to show the changes in the swarm's configuration with varying target distributions in a collision-free manner. It may be noted that when no targets are detected, the target node and assigned node can be the same, thus generating a straight-line trajectory. More detailed results with various scenarios are provided in the supplementary material.

\subsection{Performance evaluation using Monte Carlo simulation}
\begin{figure}
    \includegraphics[clip,trim=0mm 1mm 0mm 8mm,width=0.9\linewidth]{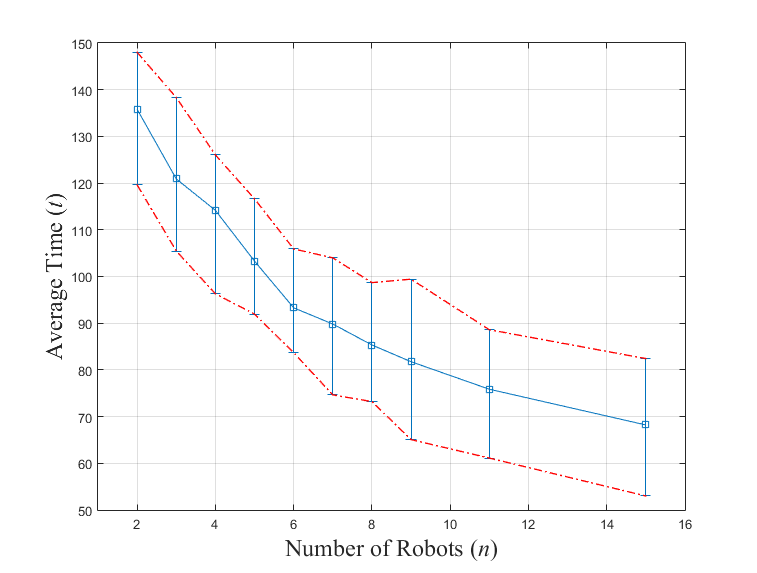}
    \caption{Results from the Monte Carlo simulation showing the average mission time.}
    \label{fig:MontoCarlo}
\end{figure}

For the Monte Carlo simulations, a search area $\mathcal{B}$  of size $200m \times 200m$ is considered with $15$ heterogeneous targets randomly spread across the area. The robot swarm is randomly initialized in the bottom left corner of  $\mathcal{B}$. The study is conducted by varying the number of robots ($3$ to $15$) in the swarm while keeping the remaining parameters constant. For a given number of robots, the simulations are repeatedly performed in this environment for $100$ times with random initialization of target locations. The time taken to detect and neutralize all the targets is measured as the performance index. The mean and standard deviation of the mission completion time is shown in fig. \ref{fig:MontoCarlo}. From the figure, it can be seen that the time taken to complete the mission decreases with an increase in the number of robots. The rate of decrease in total time decreases beyond the swarm size of $6$. It is due to the increase in the coordination times between the robots inside the swarm.
Further, the swarm size is restricted to $15$ due to the size of the search area $A$ and the coverage area of individual robots. Note that the CA-DQN algorithm runs in individual robots and is independent of the number of robots. Hence, the approach is scalable.

\begin{figure}
    \includegraphics[clip,trim=0mm 0mm 0mm 8mm,,width=0.9\linewidth]{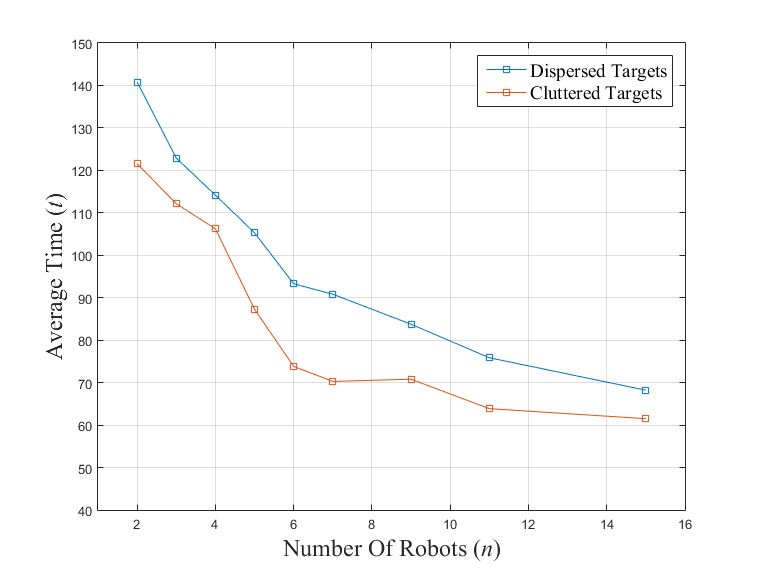}
    \caption{Results of Monte Carlo simulation for different distribution of cluttered and distributed targets.}
    \label{fig:MC_disp_scat}
\end{figure}
Another factor affecting mission performance is the target distribution. Monte-Carlo simulations (100 simulations) are conducted on two different target distributions (a uniform distribution and a clustered distribution) to study their performance effects.  Fig. \ref{fig:MC_disp_scat} shows the average time required to complete the mission (neutralize all the targets) for a varying number of robots. A cluster of targets is defined when several targets are present within a circular region of radius $\geqslant 3$. For the simulation study, it is taken to be $10$m, with a total of $15$ targets in the arena. For every mission, a random value of  $1\leqslant m \leqslant 5$ is chosen as the number of clusters. The number of targets in every cluster is taken to be $\lfloor 15/m \rfloor$. The last cluster is given the remaining targets in case of an uneven division. The centers of the clusters are randomly distributed across the arena. Between these two target distributions, CA-DQN performs better in terms of lower mission completion time for clustered targets compared to uniformly distributed targets for any given number of targets, as shown in fig. \ref{fig:MC_disp_scat}. The performance of CA-DQN among clustered targets is better, as multiple targets in a cluster would be neutralized simultaneously. However, in the case of uniformly distributed targets, the robots need significant search efforts to detect and neutralize all the targets, thus requiring more time. As the number of robots is increased, their cumulative sensing range is increased. This results in quicker detection of targets and hence early neutralization. Therefore, the mission time decreases as the number of robots is increased.

As the robots are assigned unique nodes at any given time, the algorithm is expected to avoid any collisions. A simulation study is conducted to analyze the performance of CA-DQN collision avoidance. A Monte-Carlo simulation on randomly generated 10,000 different conflicting cases is performed. The accuracy for collision avoidance is found to be $99.43\%$. The smaller percentage of collisions can be avoided by further using no motion upon approaching a minimum distance. Further, Monte Carlo studies of CA-DQN with varying robot sensing radius have been performed and are given in the supplementary material.

\subsection{Performance evaluation against coalition based approach}

\begin{figure}[]
	\centering
    \includegraphics[clip,trim=0mm 0mm 0mm 10mm,,width=0.9\linewidth]{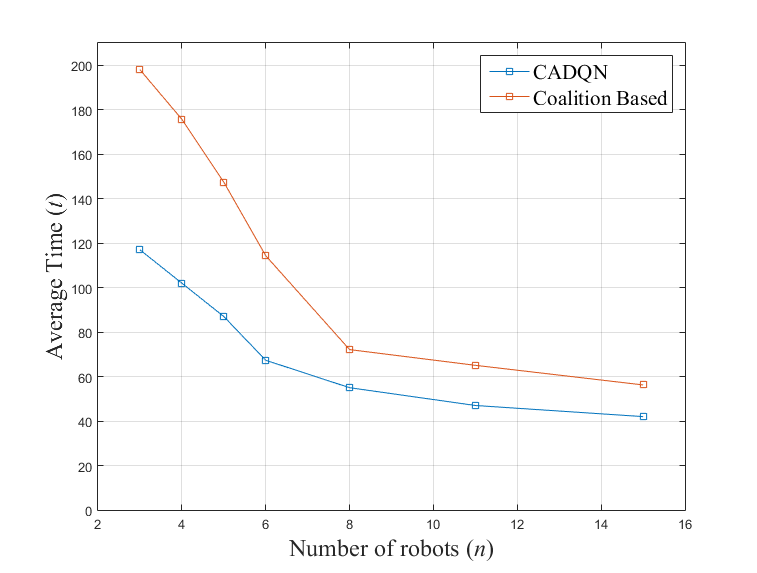}
    \caption{Average time taken by the CA-DQN and coalition based approaches to accomplish the mission}
    \label{fig:TT_comp}
\end{figure}
\begin{figure}[]
	\centering
    \includegraphics[clip,trim=0mm 0mm 0mm 8mm,,width=0.9\linewidth]{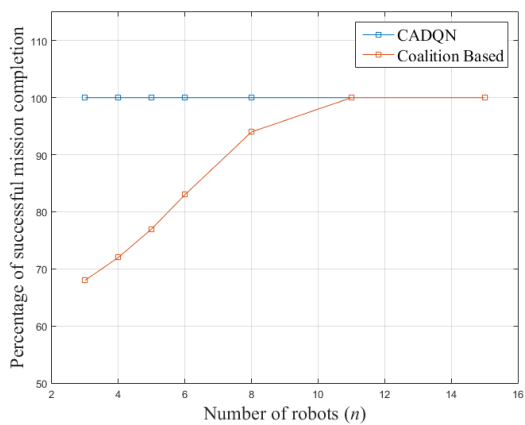}
    \caption{Comparative study of successful mission completion with increase in the number of robots.} 
    \label{fig:failure_case}
\end{figure}

It may be pointed out that \cite{Manathara2011} uses communication for a decentralized coalition formation to prosecute the targets simultaneously. However, for comparison purposes, \cite{Manathara2011} has been modified for a sequential task execution, which allows MRT neutralization. Therefore, the nearest comparison that can be made is to check the performance of CA-DQN against an approach that uses communication. The comparison study will help in understanding the efficacy of CA-DQN operating without any communication. For the comparison study, an uncertain environment of $90m \times 90 m$ is considered, and 15 targets are randomly placed in the region. Each robot is equipped with a sensor to detect targets, and the sensor radius is set to be $20m$. For CA-DQN, the swarm bounding region is set as $30m \times 30m$.   100 Monte-Carlo simulations (with each simulation time set to 300 sec) are conducted by varying the number of targets. The algorithms' performance is measured using the average simulation time and the percentage of mission success (i.e., all targets are neutralized). The results from the above study are shown in fig. \ref{fig:TT_comp} and fig. \ref{fig:failure_case}. From fig. \ref{fig:TT_comp}, it can be seen that the average time taken to complete the mission by CA-DQN is always lower than that of the coalition based approach \cite{Manathara2011}. Particularly, in cases where the swarm size is smaller than the number of targets, the difference in average simulation time is significant. The difference reduces further with the increasing number of robots, and the difference remains constant beyond $8$ robots. Fig. \ref{fig:failure_case} shows the percentage of completed missions within a specified mission time based on the Monte Carlo simulations. From the figure, it can be seen that the percentage of mission success increases with an increase in the number of robots for a coalition-based approach. In contrast, CA-DQN completes the mission effectively without any failure, within the specified simulation time with any given number of robots. The results clearly show that the swarm of robots perform search effectively using CA-DQN than the coalition-based approach presented in \cite{Manathara2011}.

\section{Experimental Study} \label{sec:4}

The proposed CA-DQN strategy has been further validated using three turtlebot3 robots in an experimental setup in the laboratory and the same is described below.  

\begin{figure}[t]
	\centering
    \includegraphics[width=0.8\linewidth]{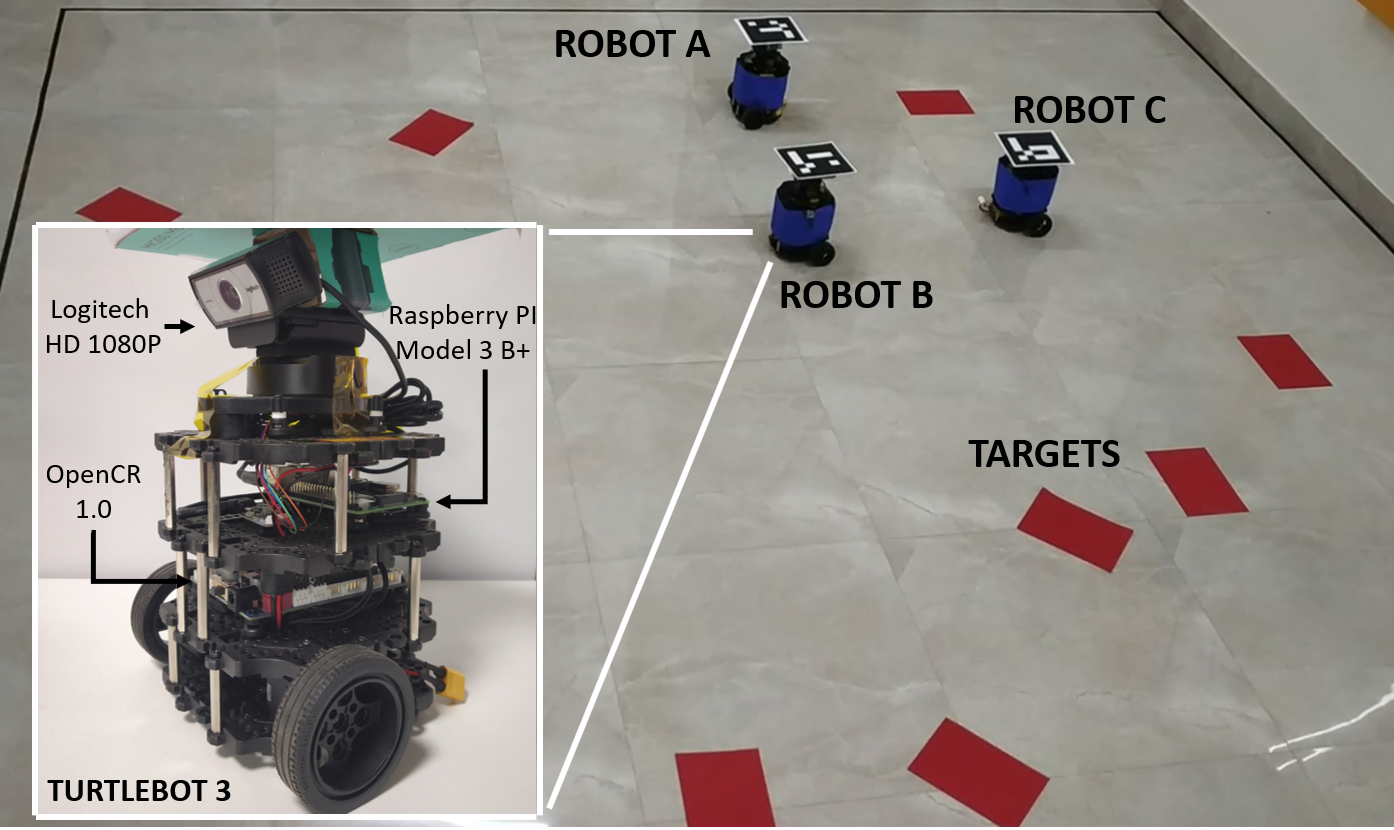}
    \caption{Experimental setup using TurtleBot3 and arena highlighting the targets.}
    \label{fig:exp_setup}
\end{figure}

\begin{figure*}
\centering
    \subcaptionbox{Simulation frame at time 26 \textit{seconds}\label{fig:a2}}
    {\includegraphics[clip,trim=0.5mm 0.5mm 0mm 0mm,width=0.33\textwidth, height=5cm]{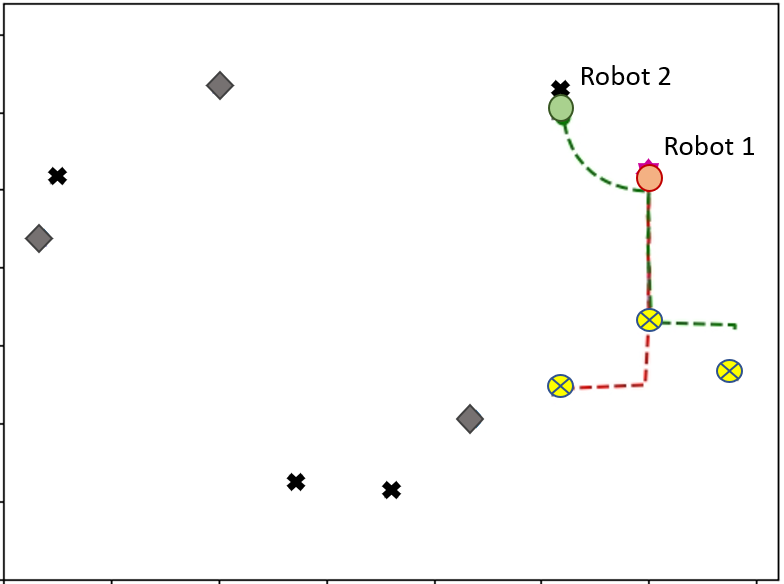}}
    \subcaptionbox{Simulation frame at time 32 \textit{seconds}\label{fig:b2}}
    {\includegraphics[clip,trim=0.5mm 0.5mm 0mm 0mm,width=0.33\textwidth , height=5 cm]{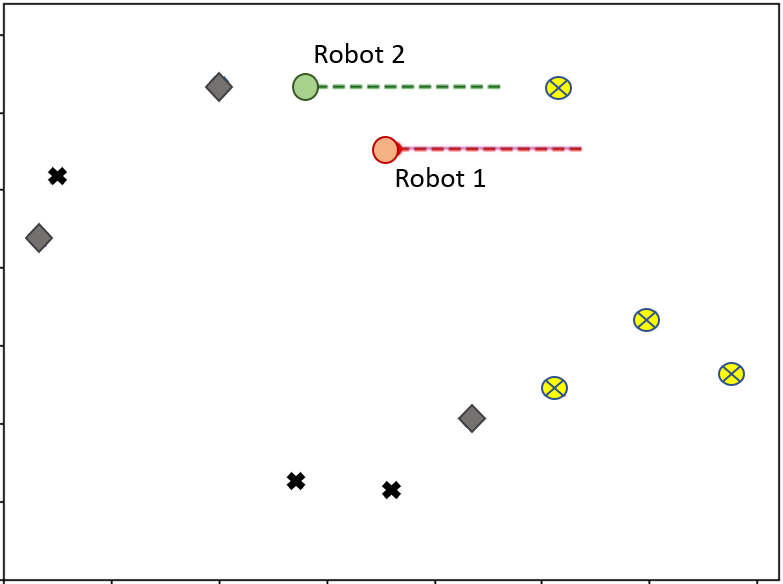}}
    \subcaptionbox{Simulation frame at time 86 \textit{seconds}\label{fig:c2}}
    {\includegraphics[clip,trim=0.5mm 0.5mm 0mm 0mm,width=0.32\textwidth , height=5 cm]{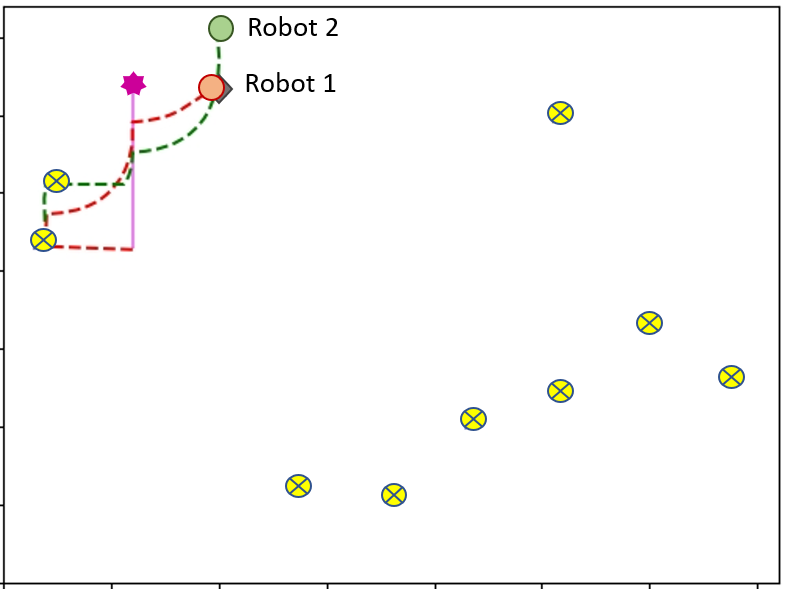}}
        \subcaptionbox{Experiment frame at time 34 \textit{seconds}\label{fig:d2}}
    {\includegraphics[clip,trim=1.2mm 1mm 0mm 0mm,width=0.33\textwidth, height=5cm]{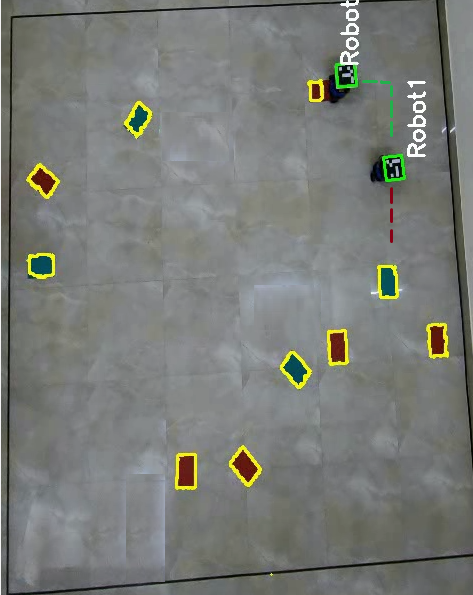}}
    \subcaptionbox{Experiment frame at time 41 \textit{seconds}\label{fig:e2}}
    {\includegraphics[clip,trim=1.2mm 1mm 0mm 0mm,width=0.33\textwidth , height=5 cm]{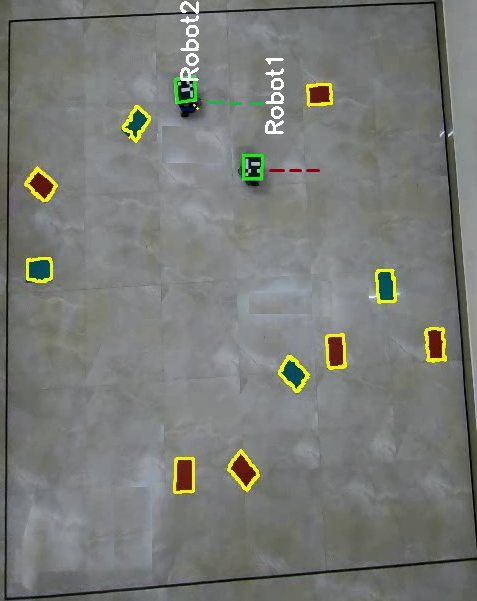}}
    \subcaptionbox{Experiment frame at time 91 \textit{seconds}\label{fig:f2}}
    {\includegraphics[clip,trim=1.2mm 1mm 0mm 0mm,width=0.32\textwidth , height=5 cm]{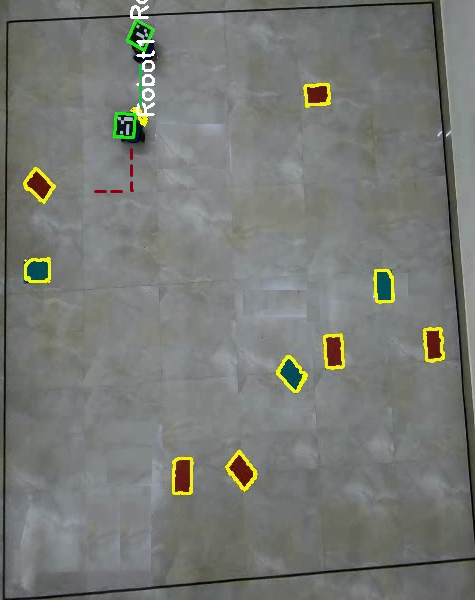}}
    \caption{The snapshots of simulation results (a-c) and experimental results (d-f) for heterogeneous targets.}
    \label{fig:sim_exp_trajectory2}
\end{figure*}

\subsection{Experimental hardware setup}

Three TurtleBot3 robots are used in the experimental setup. Each robot has a Raspberry Pi 3B+ board for onboard processing and is controlled by an OpenCR control module, which is an STM32F7 series chip with a powerful ARM Cortex-M7 processor. A Logitech C930e web-camera is placed on top of the robots as a global sensor to detect the targets, and the 360$^0$ Laser Distance Sensor LDS-01 (2D Lidar) is used as a local sensor to detect a neighbor. The experimental setup is shown in fig. \ref{fig:exp_setup}. The experiments are conducted in a laboratory environment of area $360 \times 360$ cm. A camera is mounted on the roof that monitors the robot's movements. The SRT is represented using the 'red' color, and MRT is represented using the 'green' color.  

\subsection{Experimental results for heterogeneous targets}
For experiments,  $10$ heterogeneous targets ($6$ SRTs and $4$ MRTs) are chosen and are randomly placed in the area. SRTs are marked in 'red' color, and MRTs are marked in green color, as shown in fig. \ref{fig:sim_exp_trajectory2}. In this experiment, the MRTs are assumed to require two robots to neutralize them. It also required the robots to move over the target in a sequence one after another. The robots are initialized at the bottom right corner of the arena. The robots search the area to detect the targets and go over them to neutralize. During the experimental study, the robots are equipped with a camera to detect the targets and a Lidar for detecting their neighbors. Since there is no HALE ('virtual robot') in our laboratory environment, a WiFi sensor is used to transmit the robot's position to the computer ('virtual robot'), which communicates only each robot's distance from the virtual robot.  It may be pointed out that there is no intercommunication between the robots in the swarm. The WiFi is used to mimic a virtual range sensor, and the camera is used as a HALE in the experimental setup.

The snapshots of the simulation results for heterogeneous targets are shown in fig. \ref{fig:sim_exp_trajectory2} (a-c) and the snapshots from the fish-eye camera on the experimental setup is shown in fig. \ref{fig:sim_exp_trajectory2} (d-f). There are differences between the paths followed by the robots in the experiment and the simulation. These differences are essentially due to the robot dynamics, but the sequence of allocated tasks neutralized by the robots remains the same. In Figs. \ref{fig:b2} and \ref{fig:e2}, it can be observed that the robot $R_2$ can detect the nearest target but it does not approach it to neutralize it. It is a case where the targets lie beyond the bounding radius of the swarm from the centroid; hence the robot chooses not to approach it. Similar to the previous case, one can observe the trajectory differences in Figs. \ref{fig:a2}, \ref{fig:d2} and in Figs. \ref{fig:c2} and \ref{fig:e2}. From these experimental study results, it can be concluded that the proposed CA-DQN can be implemented in a low-power computing device and is capable of handling heterogeneous targets. Further experiments with all SRT targets are given in the supplementary material.  The video of the experiment can be found in \textbf{http://bit.ly/cadqnvideo}.

\section{Conclusions} \label{sec:5}
In this paper, CA-DQN, a reinforcement learning framework for a swarm of robots to perform search and neutralize missions, has been presented. CA-DQN has been developed to perform missions as a swarm for neutralizing heterogeneous targets in an uncertain environment, with no communication among them. The CA-DQN framework uses two separate DQNs to reduce the dimension of the robot's states at any given time; as with the proposed state representation, the required size would be higher if a single network would have been used. It would lead to an increase in training time for the neural network and may affect the convergence. The convergence study of the proposed framework has been given to highlight the stable learning of the DQNs. The performance of CA-DQN has been evaluated using both simulations and real-world experiments. Detailed Monte-Carlo simulation studies by varying the number of robots, global sensing radius, and the ratio of SRTs to MRTs in the environment have been conducted. The study results indicate that CA-DQN is $100\%$ successful in completing the missions with lower average mission times than the existing state-of-the-art coalition-based approach. Since it is a decentralized approach, it is easily scalable. The results highlight the scalability and efficiency of CA-DQN while performing a reconnaissance mission. Further, the experimental study shows that CA-DQN can be implemented with low-power computing devices, which form the fundamental basis of swarm robotics. However, the current task allocation used in the paper does not guarantee a minimum time for neutralizing heterogeneous targets. Thus, finding an optimal allocation with heterogeneous targets is still an open problem. 

\bibliographystyle{IEEEtran}
\bibliography{main}

\end{document}


\maketitle

\begin{abstract}
    The supplementary material provides additional simulation and experimental setup and results for the performance of CA-DQN.
\end{abstract}

\section*{\large Working of CA-DQN}
\begin{figure*}
\centering
    \subcaptionbox{Simulation frame at  13 \textit{seconds}\label{fig1:a}}
    {\includegraphics[clip,trim=0.8mm 1mm 0mm 0mm,width=0.33\textwidth, height=5cm]{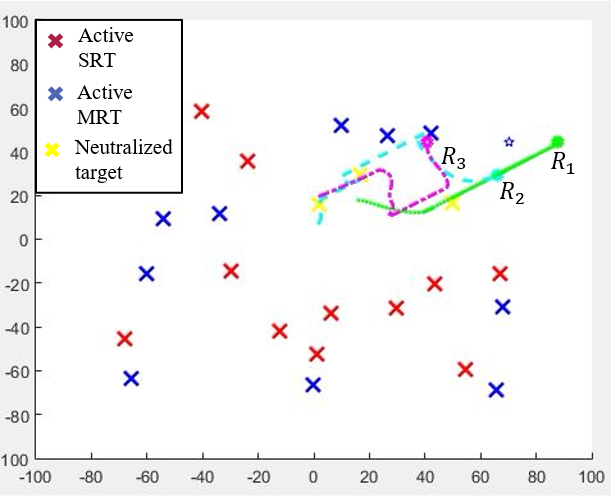}}
    \subcaptionbox{Simulation frame at  25 \textit{seconds}\label{fig1:b}}
    {\includegraphics[clip,trim=0.8mm 1mm 0mm 0mm,width=0.33\textwidth , height=5 cm]{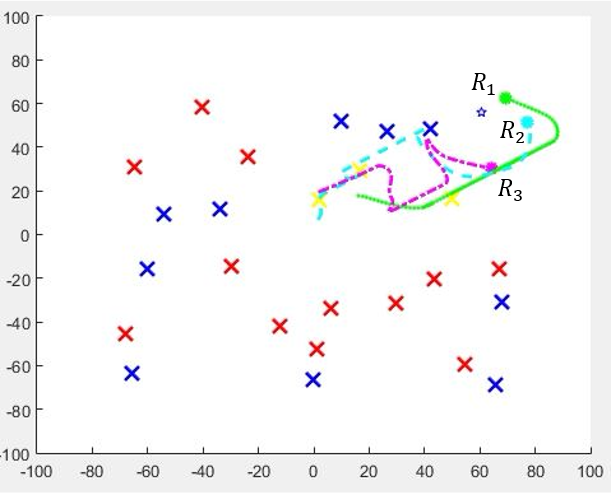}}
    \subcaptionbox{Simulation frame at  33 \textit{seconds}\label{fig1:c}}
    {\includegraphics[clip,trim=1.2mm 1mm 0mm 0mm,width=0.32\textwidth , height=5 cm]{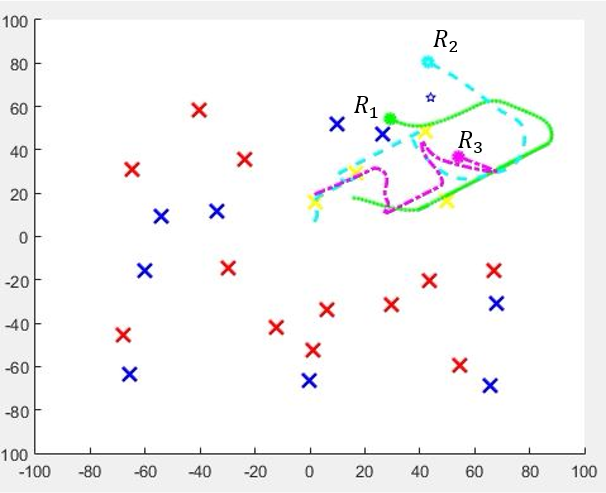}}

    \caption{The snapshots of simulation results (a-c) showing a sequence of motion of the robots to move away from environment boundary.}
    \label{fig1:sim_exp_trajectory}
\end{figure*}

To further understand the working of CA-DQN various scenarios is considered to show the flow of information. In \ref{fig1:sim_exp_trajectory}, a sequence of robot motions are shown to show the movement of robots when approaching the environment boundary. AS mentioned earlier, the robots mask the grid nodes that are beyond this boundary. Therefore the robots get deflected from the boundary by randomly choosing one of the remaining nodes. It may be noted that other robots start deflecting even before reaching the boundary. It happens because, as the first robot moves away from the boundary, the centroid also shifts accordingly.

\begin{figure*}
\centering
    \subcaptionbox{Simulation frame at  45 \textit{seconds}\label{fig2:a}}
    {\includegraphics[clip,trim=0.8mm 1mm 0mm 0mm,width=0.33\textwidth, height=5cm]{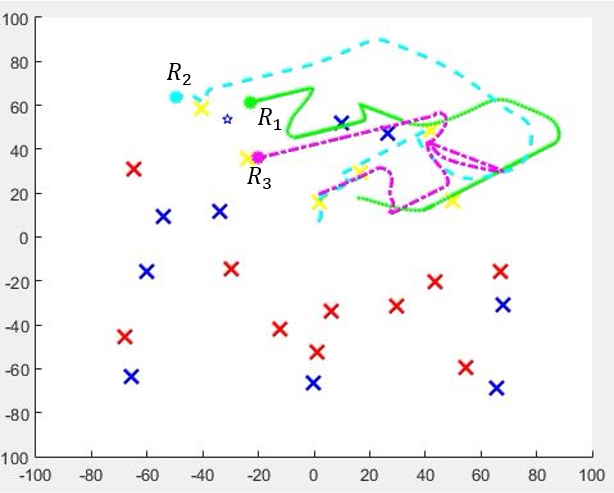}}
    \subcaptionbox{Simulation frame at  56 \textit{seconds}\label{fig2:b}}
    {\includegraphics[clip,trim=0.8mm 1mm 0mm 0mm,width=0.33\textwidth , height=5 cm]{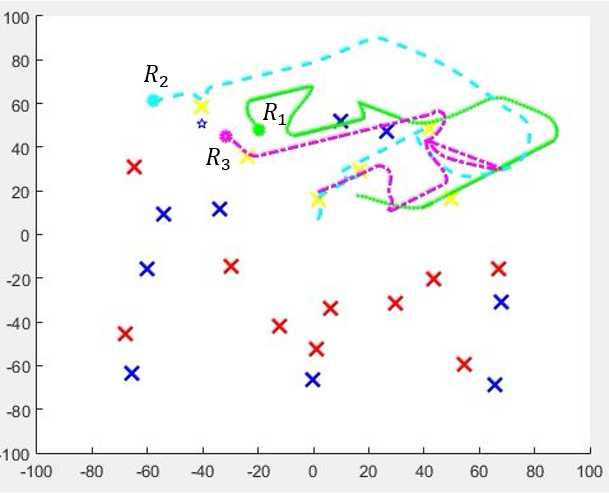}}
    \subcaptionbox{Simulation frame at  63 \textit{seconds}\label{fig2:c}}
    {\includegraphics[clip,trim=1.2mm 1mm 0mm 0mm,width=0.32\textwidth , height=5 cm]{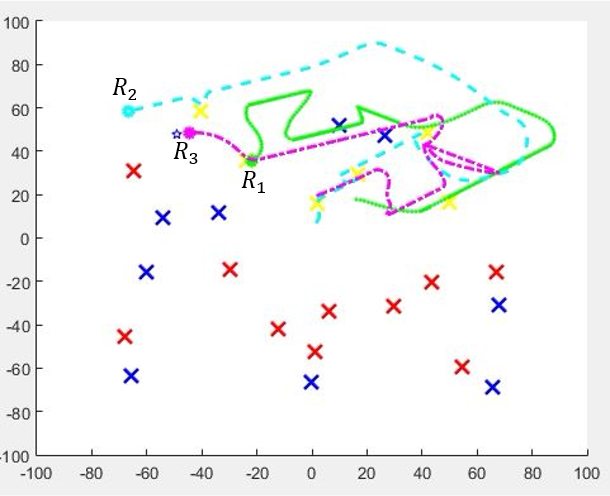}}

    \caption{The snapshots of simulation results (a-c) showing a sequence of motion of the robots to ensure maximum search area within the swarm.}
    \label{fig2:sim_exp_trajectory}
\end{figure*}

Fig. \ref{fig2:sim_exp_trajectory} shows the sequence of motion when the robots detect no target and attempt to maximize the search area coverage within the swarm. In fig. \ref{fig2:a} $R_3$ neutralizes a target and then there is no active target left within the sensing region. Thus the robots now aim to maximize spread. As $R_3$ is already ahead, it moves further ahead, towards the center, as $R_1$ starts coming back to the left away from $R_2$ and $R_3$ simultaneously. The final spread can be seen in fig. \ref{fig2:b} with the robots separated from each other with sensor radius of each being $15m$.

\begin{figure}
\centering
    \subcaptionbox{Simulation frame at  133 \textit{seconds}\label{fig4:a}}
    {\includegraphics[clip,trim=0mm 1mm 0mm 0mm,width=0.33\textwidth, height=5cm]{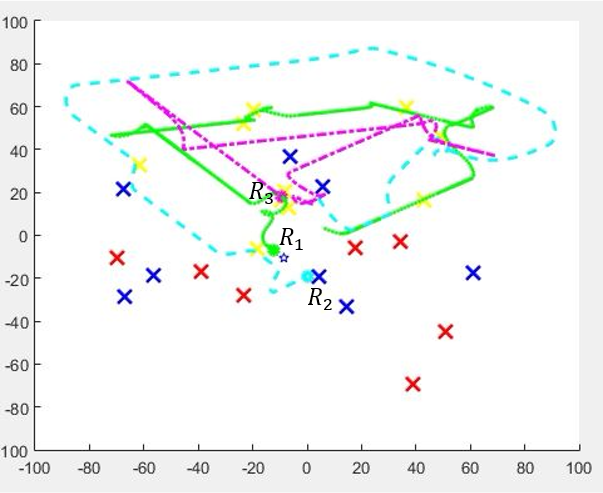}}
    \subcaptionbox{Simulation frame at  149 \textit{seconds}\label{fig4:b}}
    {\includegraphics[clip,trim=0mm 1mm 0mm 0mm,width=0.33\textwidth , height=5 cm]{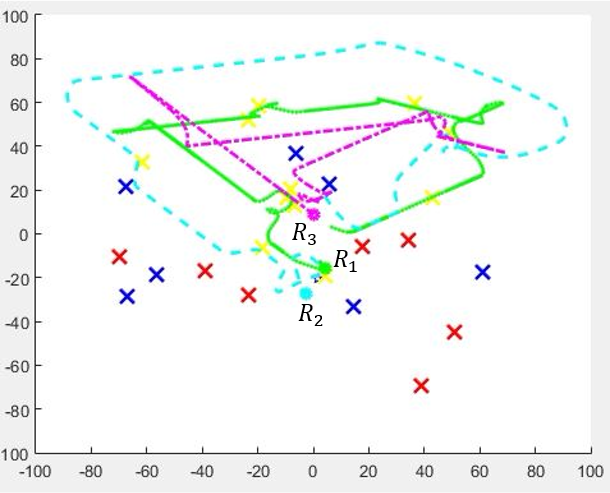}}
    \subcaptionbox{Simulation frame at  161 \textit{seconds}\label{fig4:c}}
    {\includegraphics[clip,trim=0mm 1mm 0mm 0mm,width=0.33\textwidth , height=5 cm]{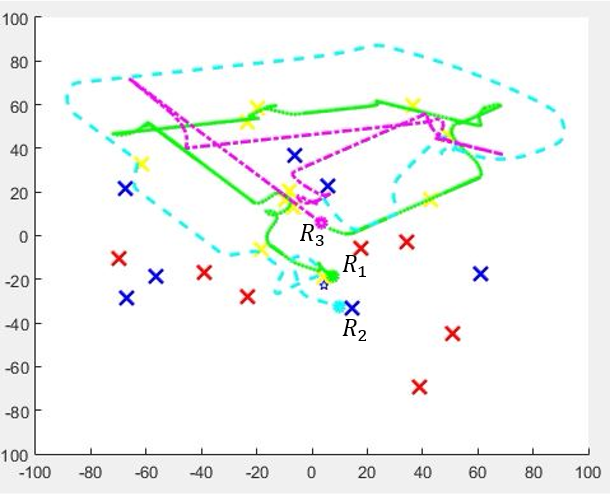}}
    \subcaptionbox{Simulation frame at  173 \textit{seconds}\label{fig4:d}}
    {\includegraphics[clip,trim=0mm 1mm 0mm 0mm,width=0.33\textwidth, height=5 cm]{graphics/p4_3.png}}
    

    \caption{The snapshots of simulation results (a-c) showing a sequence of motions of the robot to allocate the targets and neutraliza a MRT.}
    \label{fig3:sim_exp_trajectory}
\end{figure}

Fig. \ref{fig3:sim_exp_trajectory} show the sequence of robot motion when the robots cooperate to neutralize MRTs. In fig. \ref{fig4:a} $R_2$ takes a target. Then if fig. \ref{fig4:b} it can be seen that $R_2$ makes space for $R_1$ to attack on the MRT to neutralize it. Following that, as $R_2$ is closer to the next MRT, the targets are then $R_1$. After $R_2$ attacks the MRT in fig. \ref{fig4:c},similar to that in the previous case, it creates space for the $R_1$ to attack as shown in fig.\ref{fig4:d}. This shows the robots to cooperatively neutralize a MRT without communication with each other under the CA-DQN framework.

\section*{\large Performance Evaluation Using Monte Carlo Simulation}

\begin{figure}[t]
    \includegraphics[clip,trim=0mm 3mm 0mm 8mm,width=0.95\linewidth]{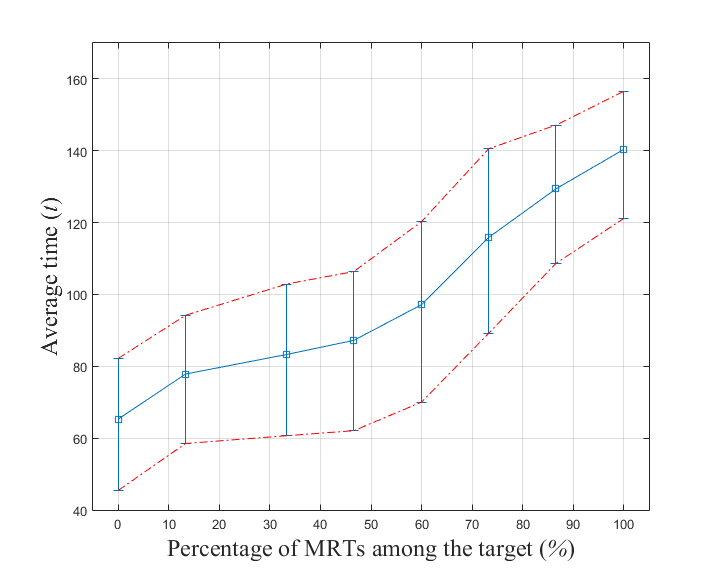}
    \caption{Monte Carlo simulation results with varying percentage of MRTs among the total number of targets.}
    \label{fig:MontoCarlo2}
\end{figure}

As mentioned earlier in the paper, for the Monte Carlo simulations, a search area $\mathcal{B}$ of size $90m \times 90m$ is considered with $15$ heterogeneous targets randomly spread across the area. The context grid for each robot is of size $5 \times 5$. The robot swarm is randomly initialized in the bottom left corner of $\mathcal{B}$. The study was conducted by varying the 
percentage of MRTs among the targets ($0\%$ to $100\%$) while keeping the rest of the parameters as constant. 
The number of robots in the swarm is taken to be $6$. 
The simulations were performed in this environment repeatedly for $50$ times with random initialization of target locations.
The mean and standard deviation of time taken to complete the mission is shown in Fig. \ref{fig:MontoCarlo2}.
From the figure, we can see that the time taken to complete the mission increases with an increase in the percentage of MRTs.

\begin{figure}
    \includegraphics[clip,trim=0mm 0mm 0mm 8mm,,width=0.95\linewidth]{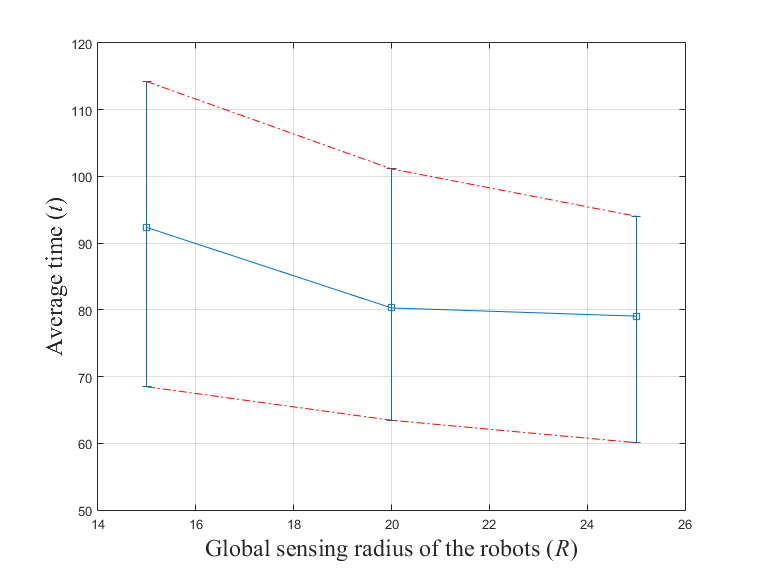}
    \caption{Monte Carlo simulation results with varying global sensing radius of the constituent robots in the swarm.}
    \label{fig:MontoCarlo3}
\end{figure}

Different sensor radii ranging from $15$m to $25$m were considered, to study the impact of the change in global sensing radius. The average time taken for the robots in the swarm to neutralize all the targets is shown in Fig. \ref{fig:MontoCarlo3}. From the figure, we can see that the time is taken decreases with an increase in the global sensor radius. However, the rate of decrease is very less beyond $20m$ radius. It could be due to high overlapping information with an increase in the sensing range beyond $20m$. 

\section*{\large Experimental Setup: Software stack and integration of CA-DQN}

The experiments carried out to validate the results of CA-DQN were conducted using ROS based TurtleBot3 ground robots in an area $360cm \times 360cm$ in the laboratory. A fixed camera was mounted at the roof of the lab to monitor the robot movements. Three Turtlebot3 robots were used, each of which was equipped with a Raspberry Pi 3B+ on-board computer. The Raspberry Pi houses a 1.4 GHz, Broadcom BCM2837B0, Cortex-A53 (ARMv8) 64-bit CPU with a 1 GB LPDDR2 memory and runs on Raspbian Stretch Linux based operating system. OpenCR (Open-source Control module for ROS) is the main control board on the TurtleBot3 robot. It houses a 216 Mhz, STM32F746ZGT6 / 32-bit ARM Cortex®-M7 controller. It also has a three-axis gyroscope, a three-axis accelerometer, and a three-axis Magnetometer (MPU9250), which is used for the robot's localization. A Logitech C930e web-camera is placed on top of the robots as a global sensor to detect the targets and the 360$^0$ Laser Distance Sensor LDS-01 (2D Lidar) is used as a local sensor to detect a neighbor.

\begin{figure}
	\centering
    \includegraphics[width=0.8\linewidth]{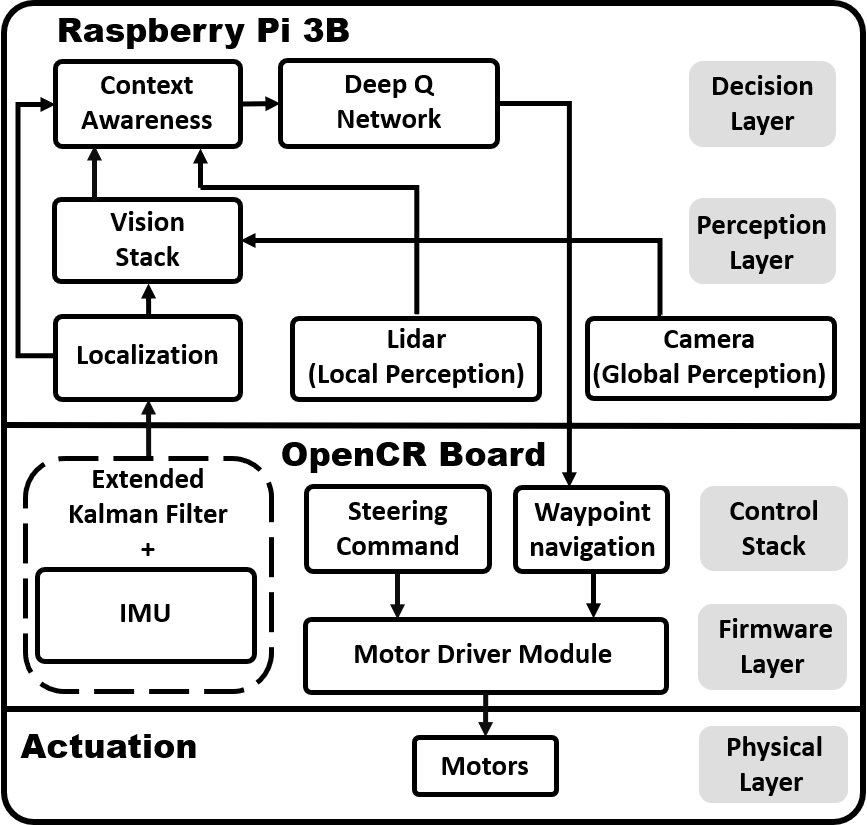}
    \caption{System architecture for implementation of CA-DQN in TurtleBot3.}
    \label{fig:architecture}
\end{figure}

\begin{figure}
	\centering
    \includegraphics[width=1\linewidth]{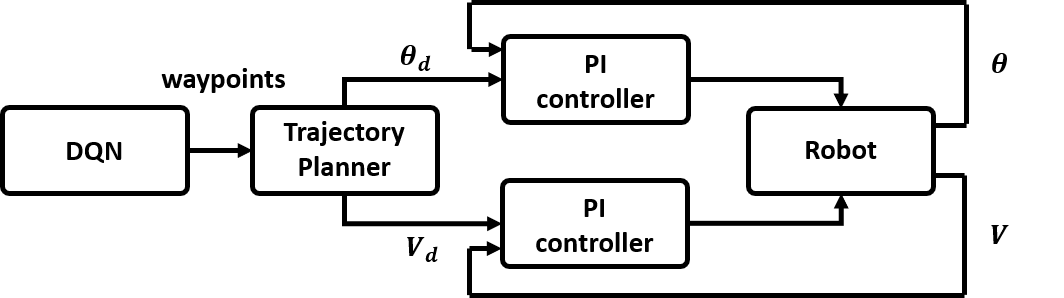}
    \caption{Control system design for  implementation of CA-DQN in TurtleBot3.}
    \label{fig:controller}
\end{figure}

Fig. \ref{fig:architecture} shows the system architecture implemented in the TurtleBot3 Burger robot. The control layer realizes the steering control and trajectory tracking control using the PI controller \cite{turtlebot_navigation}. The control and firmware layers are realized on the OpenCR ARM Cortex-M7 board. The sensor layer, perception layer, decision control layer, and navigation layer are realized on a Raspberry Pi 3B+ board using ROS. 
Fig. \ref{fig:controller} shows the control system design of CA-DQN implementation. The output given by the conflict/conflict-free DQNs provides a robot with a waypoint to follow. The desired angular position $\theta$ and desired velocity $v_d$ are used as reference PI controllers. Based on the robot's measurement feedback, error calculation is done, and the PI controller responds accordingly.

\begin{figure*}
\centering
    \subcaptionbox{Simulation frame at  22 \textit{seconds}\label{fig:a}}
    {\includegraphics[clip,trim=0.8mm 1mm 0mm 0mm,width=0.33\textwidth, height=5cm]{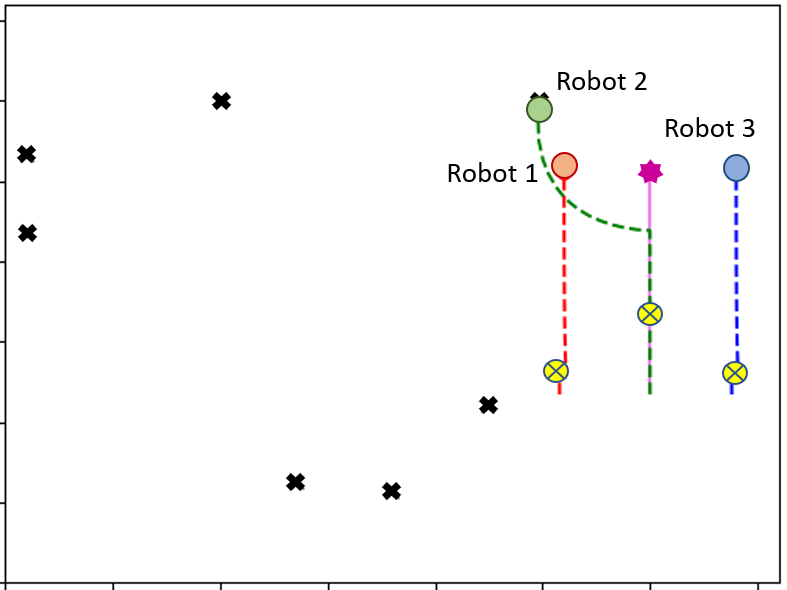}}
    \subcaptionbox{Simulation frame at  45 \textit{seconds}\label{fig:b}}
    {\includegraphics[clip,trim=0.8mm 1mm 0mm 0mm,width=0.33\textwidth , height=5 cm]{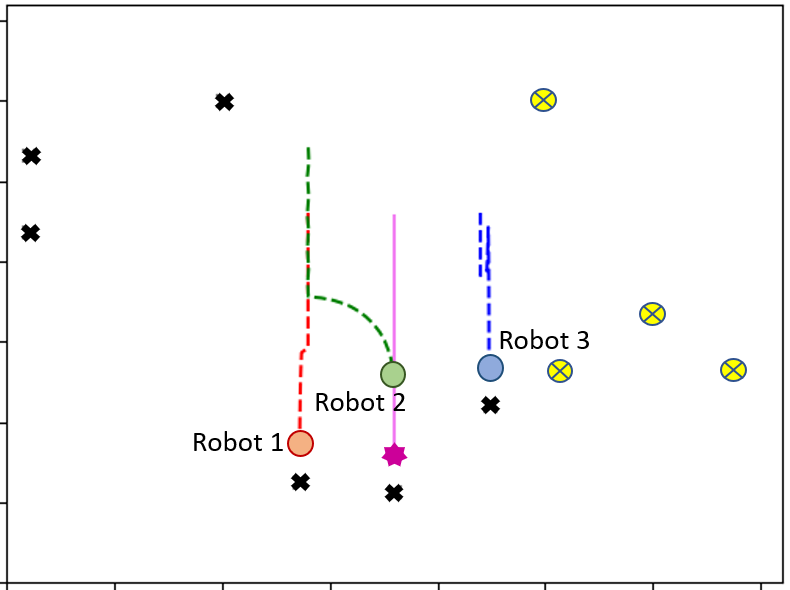}}
    \subcaptionbox{Simulation frame at  73 \textit{seconds}\label{fig:c}}
    {\includegraphics[clip,trim=1.2mm 1mm 0mm 0mm,width=0.32\textwidth , height=5 cm]{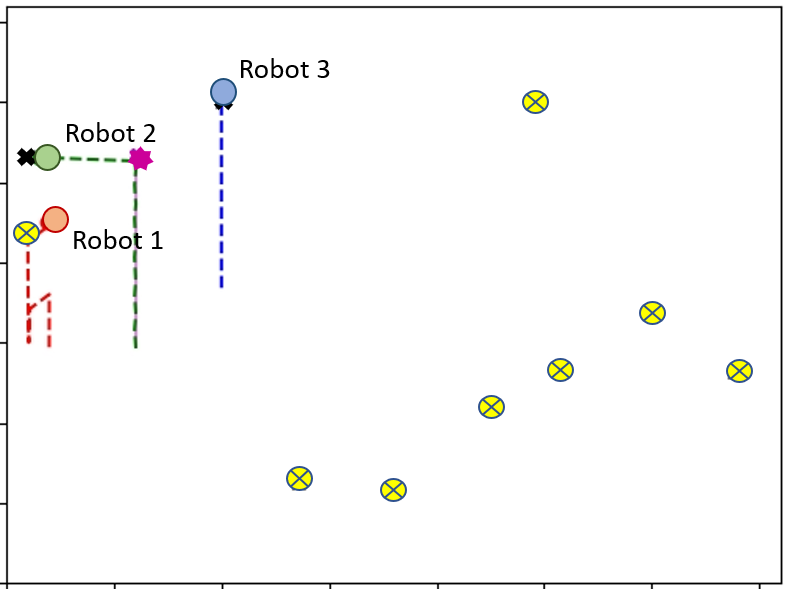}}
        \subcaptionbox{Experiment frame at  31 \textit{seconds} \label{fig:d}}
    {\includegraphics[clip,trim=1.2mm 1mm 0mm 0mm,width=0.33\textwidth, height=5cm]{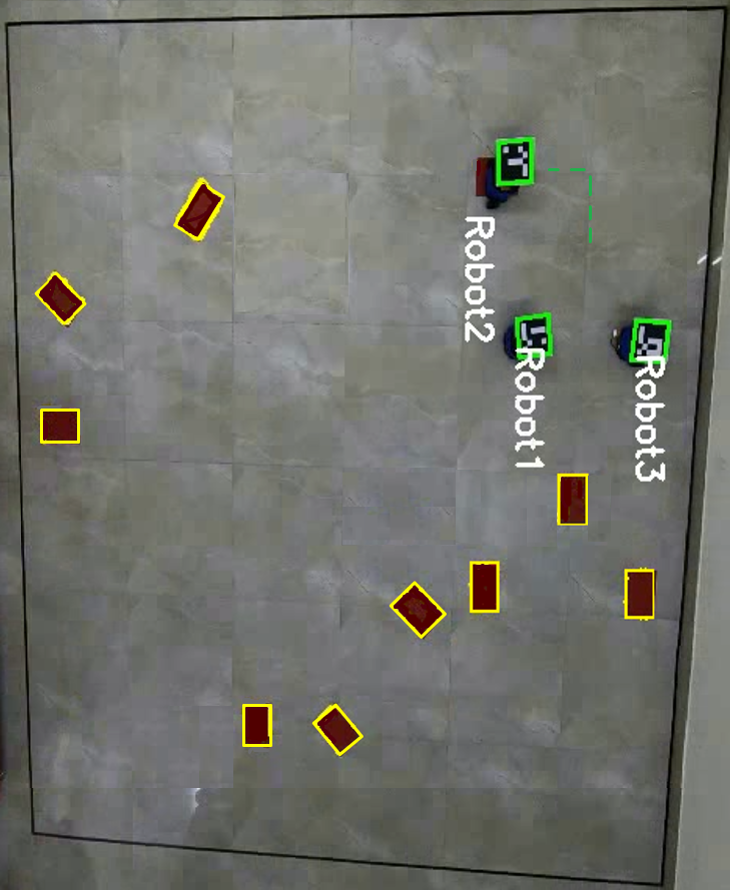}}
    \subcaptionbox{Experiment frame at  48 \textit{seconds}\label{fig:e}}
    {\includegraphics[clip,trim=1.2mm 1mm 0mm 0mm,width=0.33\textwidth , height=5 cm]{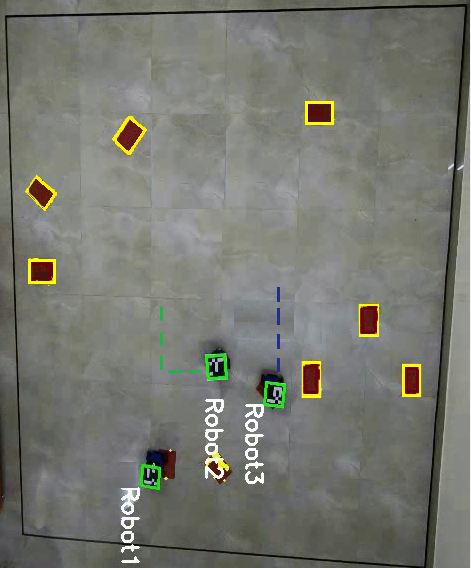}}
    \subcaptionbox{Experiment frame at  79 \textit{seconds}\label{fig:f}}
    {\includegraphics[clip,trim=1.2mm 1mm 0mm 0mm,width=0.32\textwidth , height=5 cm]{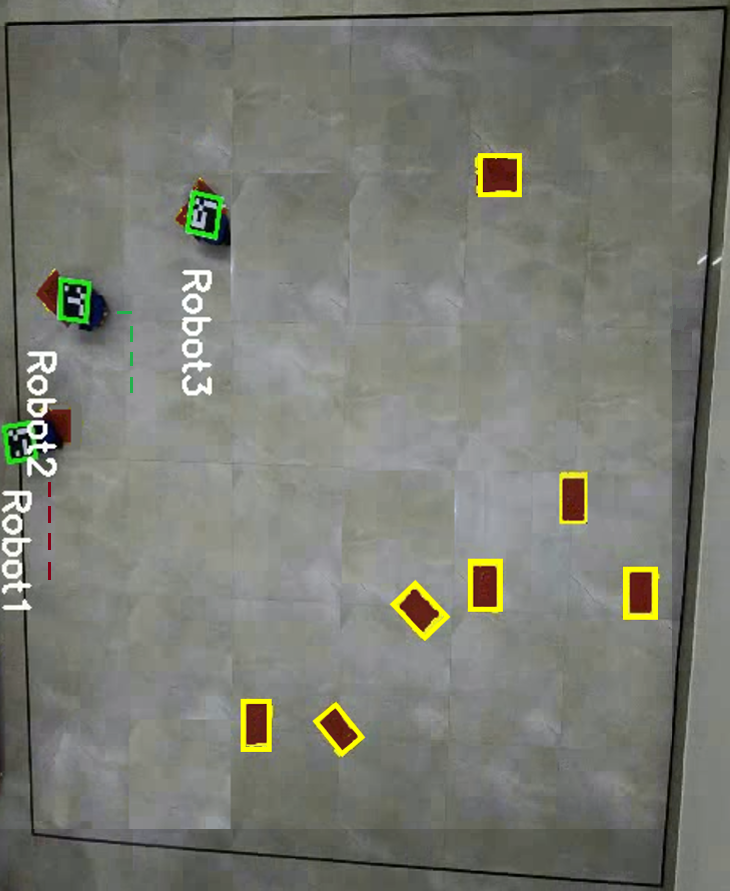}}
    \caption{The snapshots of simulation results (a-c) and experimental results captured by the fish eye camera (d-f) for SRT targets.}
    \label{fig:sim_exp_trajectory}
\end{figure*}

\begin{figure}
	\centering
    \includegraphics[clip,trim=0mm 4mm 0mm 8mm,width=1\linewidth]{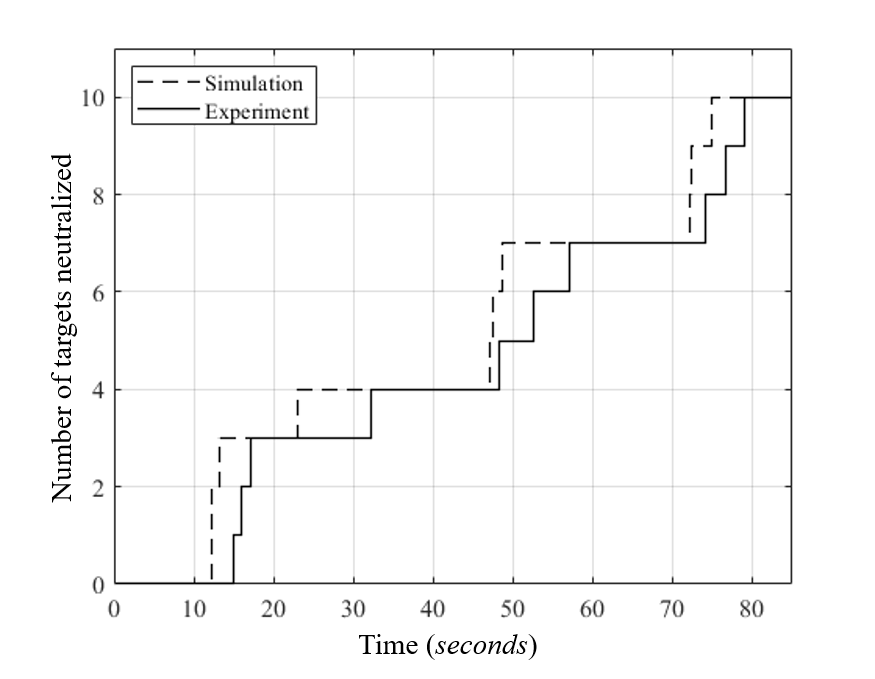}
    \caption{The time history of target neutralization from simulation and experimental setup for homogeneous targets.}
    \label{fig:case1}
\end{figure}

\begin{figure}
	\centering
    \includegraphics[clip,trim=0mm 4mm 0mm 8mm,width=1\linewidth]{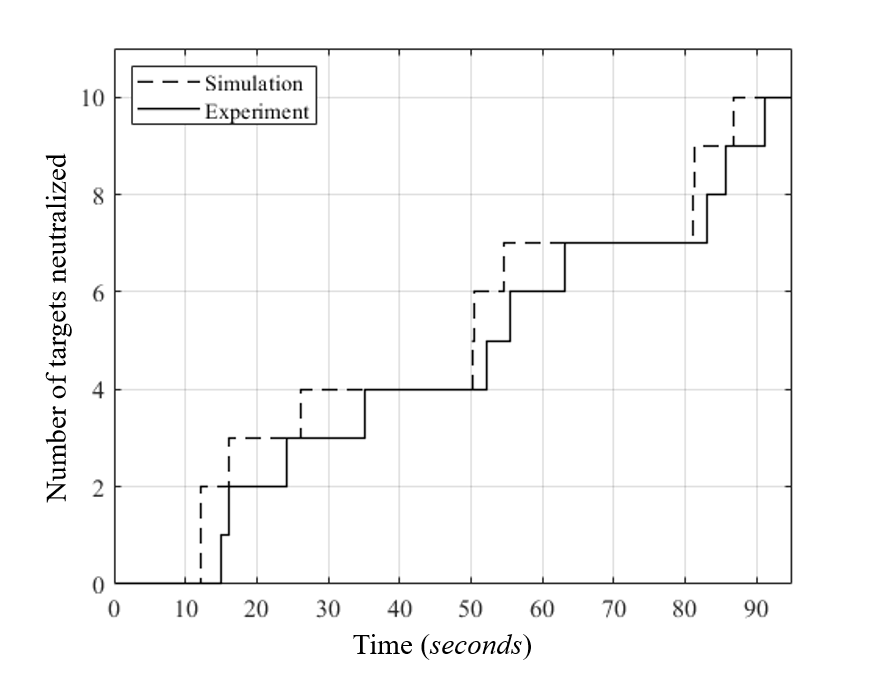}
    \caption{The time history of target neutralization from simulation and experimental setup for heterogeneous targets.}
    \label{fig:case2}
\end{figure}

\begin{figure*}
\centering
    \subcaptionbox{Experiment frame at 37 \textit{seconds}\label{fig:ea}}
    {\includegraphics[clip,trim=0.8mm 1mm 0mm 0mm,width=0.33\textwidth, height=5cm]{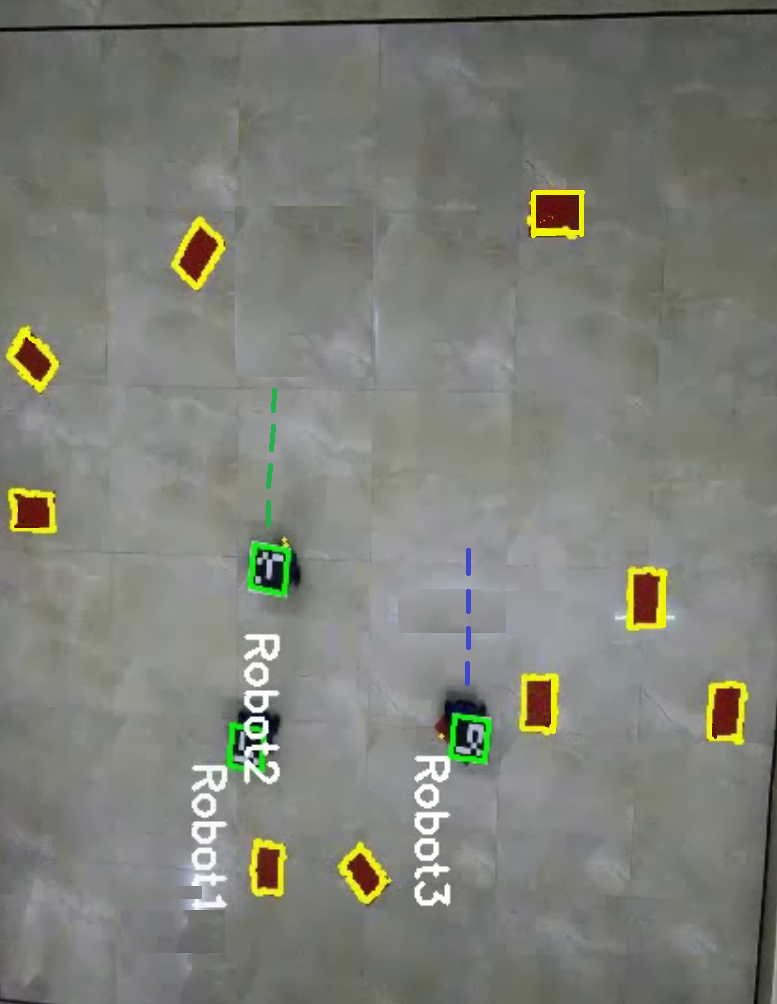}}
    \subcaptionbox{Experiment frame at 42 \textit{seconds}\label{fig:eb}}
    {\includegraphics[clip,trim=0.8mm 1mm 0mm 0mm,width=0.33\textwidth , height=5 cm]{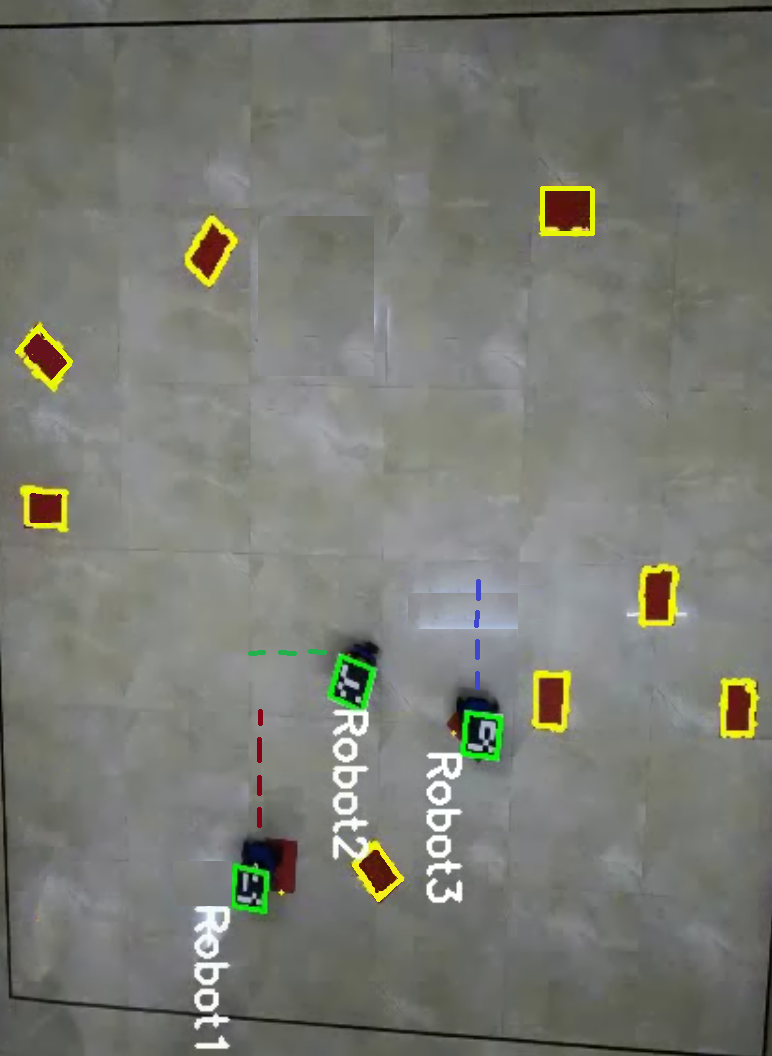}}
    \subcaptionbox{Experiment frame at 46 \textit{seconds}\label{fig:ec}}
    {\includegraphics[clip,trim=1.2mm 1mm 0mm 0mm,width=0.32\textwidth , height=5 cm]{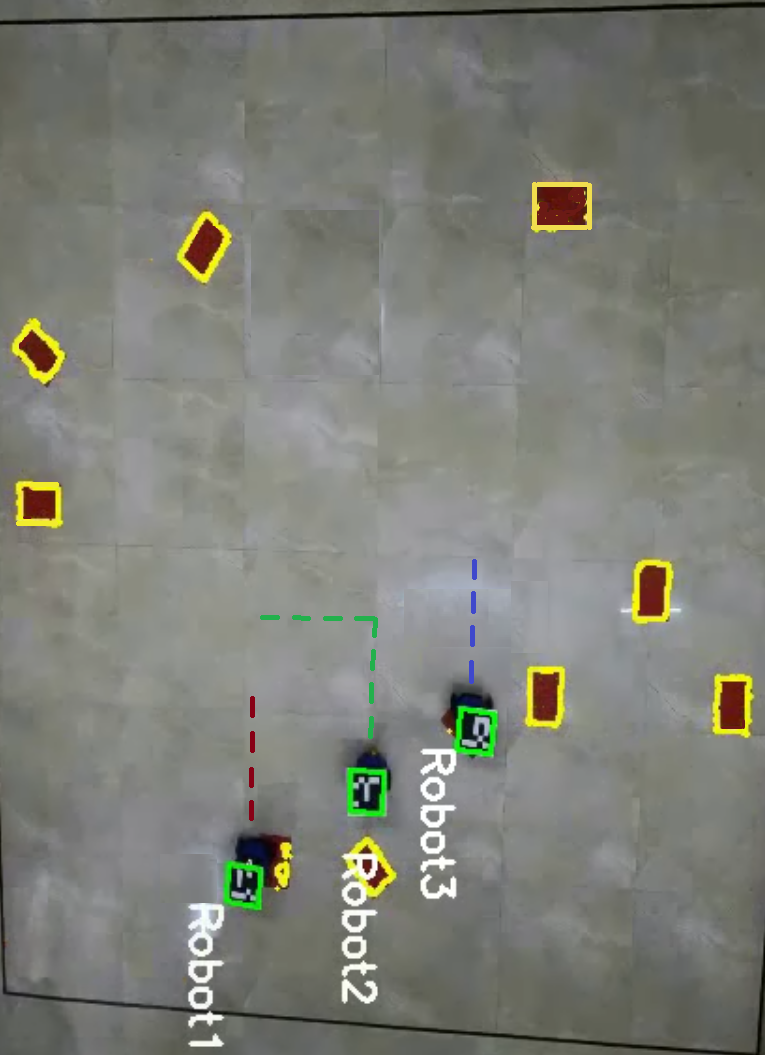}}
    \caption{The snapshots of experimental results captured by the fish eye camera (d-f) for SRT targets, showing multiple stages of reconfiguration of the swarm.}
    \label{fig:exp_trajectory}
\end{figure*}

The on-board IMU is fused with the Extended Kalman Filtering algorithm to provide local positioning of the robot. A generic ROS based robot localization package is used \cite{robot_localization}. The odometry data is used for the realization of the robot in the local grid\cite{robot_pose_ekf}. A ROS driver for V4L USB cameras \cite{usb_cam} was used to obtain the camera feed from Logitech C930e webcam, which is sent to the vision stack for color detection algorithm and reference image transformation \cite{opencv_apps} to detect targets and its position for the local grid for the robots to move. The data from the LDS-01 (2D Lidar) sensor is used to detect the neighbours\cite{turtlebot3_slam}, and while using the way-points \cite{follow_waypoints}, navigation layer realizes point-to-point navigation and provides the necessary commands to the control layer. The motor driver module executes the commands received from the control layer to the physical layer i.e., the motor actuation.

\section*{\large Experimental Results: Homogeneous Targets}

For the experiment, $10$ targets (only SRT) are placed in the arena and are marked in 'red' color, as shown in Fig. \ref{fig:sim_exp_trajectory}. The robots are initialized at the bottom right corner of the arena. The robots search the area to detect the targets. The robots are required to go over the targets to neutralize them. 
As mentioned in the paper earlier, for experimental study in the lab, the robots were equipped with a vision camera to detect the targets and a Lidar sensor for detecting the neighbors. Since there is no HALE ('virtual robot') in our laboratory environment, we have used a WiFi sensor to transmit the robot's position to the computer ('virtual robot'), which communicates the distance from the virtual robot only.
This helps the swarm of robots to stay within the bounding radius of the swarm. 
The simulation and their respective experimental frames in certain scenarios are shown in Fig. \ref{fig:sim_exp_trajectory} (a-f). Although the allocated target remains the same, differences in the path taken by the robots in simulation and corresponding experiment scenarios can be observed. These differences are a result of varying model of the robot in the simulation and experiment. This is also due to uncertainties in the experimental setup like wheel-slip and the difference in motor rpm.
The path traveled by the robots in the experiment and simulation is highlighted for better understanding. Fig. \ref{fig:a} and Fig. \ref{fig:d} show the simulation and experimental results of a scenario in which robot 2 takes a rectangular trajectory, whereas it is a smooth trajectory in the simulation. Similarly, in Fig. \ref{fig:b} and Fig. \ref{fig:c}, the robot three and robot one respectively are taking paths opposite to swarm and correcting themselves according to the assigned node, which is not emulated in the experiment. 
Fig. \ref{fig:case1} and \ref{fig:case2} show the time instances for the targets to neutralize as a square plot for simulation and the experiment in various cases like homogeneous and heterogeneous targets respectively.

Consider the case shown in Fig. \ref{fig:e}; to achieve this configuration for neutralizing the detected targets, the swarm undertook multiple stages of reconfiguration, as shown in Fig.\ref{fig:exp_trajectory}. Upon detection of targets, the robots are assigned to the respective targets. Hence, the robots use CA-DQN to change their configuration as soon as possible to neutralize the targets.

\bibliographystyle{IEEEtran}
\bibliography{supp}